\documentclass[3p]{elsarticle}

\usepackage[T1]{fontenc}
\usepackage[utf8]{inputenc}
\usepackage[english]{babel}

\usepackage{hyperref}

\begin{document}

\title{Spin Insulatronics}

\author[1]{Arne Brataas}

\address[1]{Center for Quantum Spintronics, Department of Physics, Norwegian University of Science and Technology, NO-7491 Trondheim, Norway.}%
\ead{Arne.Brataas@ntnu.no}

\author[2]{Bart van Wees}
\address[2]{Physics of Nanodevices, Zernike Institute for Advanced Materials, University of Groningen, Nijenborgh 4, 9747 AG Groningen, The Netherlands}

\author[3]{Olivier Klein}
\address[3]{SPINTEC, CEA-Grenoble,CNRS and Université Grenoble Alpes, 38054 Grenoble, France}

\author[4]{Gr\'{e}goire de Loubens}
\author[4]{Michel Viret}
\address[4]{SPEC, CEA-Saclay, CNRS, Université Paris-Saclay, 91191 Gif-sur-Yvette, France
}

\date{\today}% It is always \today, today,
             %  but any date may be explicitly specified

\begin{abstract}
Spin insulatronics covers efforts to generate, detect, control, and utilize high-fidelity pure spin currents and excitations inside magnetic insulators. Ultimately, the new findings may open doors for pure spin-based information and communication technologies. The aim is to replace moving charges with dynamical entities that utilize low-dissipation coherent and incoherent spin excitations in antiferromagnetic and ferromagnetic insulators. The ambition is that the new pure spin-based system will suffer reduced energy losses and operate at high frequencies. In magnetic insulators, there are no mobile charge carriers that can dissipate energy. Integration with conventional electronics is possible via interface exchange interactions and spin-orbit couplings. In this way, the free electrons in the metals couple to the localized spins in the magnetic insulators. In turn, these links facilitate spin-transfer torques and spin-orbit torques across metal-insulator interfaces and the associated phenomena of spin-pumping and charge-pumping. The interface couplings also connect the electron motion inside the metals with the spin fluctuations inside the magnetic insulators. These features imply that the system can enable unprecedented control of correlations resulting from the electron-magnon interactions. We review recent developments to realize electric and thermal generation, manipulation, detection, and control of pure spin information in insulators.
\end{abstract}

%\keywords{Suggested keywords}%Use showkeys class option if keyword
                              %display desired
\maketitle

\tableofcontents

\section{\label{sec:introduction}Introduction}

Spintronics has proven its worth by causing a revolution in data storage \cite{1fert}. Nevertheless, the majority of spintronics devices continue to function via mobile electrons, which inherently dissipate power due to resistive losses. Propagating excitations of localized magnetic moments can also carry spin currents. However, in metals, spin excitations strongly attenuate due to considerable viscous damping. Only in magnetic insulators do spin currents propagate with significantly reduced dissipation since there are no conduction electrons dissipating heat. The coined term {\it spin insulatronics} includes efforts to transport spin information without electronic carrier transport. Additionally, the clear separation between spin dynamics and electron motion in hybrid systems of magnetic insulators and metals (semiconductors) can induce new states of matter. These new ways of utilizing the spin in magnetic insulators can be of a fundamental interest and pave the way for new devices. The aim is to facilitate a revolution in information and communication technologies by controlling electric signals through the deployment of antiferromagnetic insulators (AFIs) and ferromagnetic insulators (FIs). For spin insulatronics to succeed, spin signals in magnetic insulators must seamlessly integrate with conventional electronics, which is the only way the manipulation of a charge signal in an insulator can become feasible and useful in devices. 

Experimental and theoretical studies of spin dynamics in ferromagnetic and antiferromagnetic insulators have a long history \cite{Gurevich:CRC96}. The field, however, has been living through a rapid revival and excitement in recent years due to the fabrication advances in nanostructured hybrid structures of magnetic insulators and normal metals. It was reported that currents in metals can be converted to spin currents therein that in turn can surprisingly efficiently pass across the metal-insulator interface \cite{Kajiwara:nat2010}. Inside the magnetic insulators, the exceptional low spin memory loss facilitates new ways of long-range spin transport and manipulation. Spin transport and dynamics in magnetic insulator heterostructures are fundamentally different from its counterpart in metals and semiconductors. Significant progress in understanding the similarities and differences as compared to conventional spin transport is emerging. 

Spin insulatronics offers several novelties. The small energy losses in insulators enable transport of spin information across distances of tens of microns  \cite{2cornelissen,3lebrun}, much farther than in metals. Furthermore, magnetic insulators can transfer and receive spin information in new ways with respect to metals \cite{Heinrich:PRL2011,121jungfleisch,84xiao,80kapelrud,87hamadeh,85du,86demidov}. Since the transport is anisotropic, such devices can also be controlled by using an external field to change the magnetic configuration. In magnetic insulators, it is possible to realize magnon-based transistors \cite{116cornelissen}. One can also vision the electrical control of magnon-based majority gates \cite{Fischer:APL2017}. In other words, critical logical operations can be carried out by spin excitations without mobile electrons.

The reduced dissipation also facilitates quantum coherent phenomena. Magnons are bosons and can condense  \cite{4demokritov}. Condensation occurs when a macroscopic fraction of the bosons occupy the lowest energy state and when all these states are phase-coherent. Normally, at equilibrium, condensation occurs below a critical temperature set by the density of the bosons. In driven systems, like parametric pumping of magnons, condensation is an out-of-equilibrium phenomenon that sets in when the external perturbation is strong enough. One can achieve the state of spin superfluidity  \cite{5halperin}, an entirely new route to mediate spin information \cite{95bender,136sonin,137sonin,138takei,142qaiuzadeh}. A boson condensate has a fixed phase. When the phase varies slowly along a spatial dimension, the system drives a supercurrent that is proportional to the gradient of the phase. In the case of magnons, the supercurrent is a pure spin current. The system can exhibit spin superfluidity. Spin fluctuations such as magnons can also induce new properties in adjacent metals or semiconductors. An exciting possibility is magnon-induced superconductivity \cite{Kargarian:PRL2016,Gong:ScienceAdvances:2017,Rohling:PRB2018}. Magnons can replace phonons that cause electron pairing in low-temperature superconductors. This replacement opens news ways of controlling unconventional forms of superconductivity. Another possibility is to use spin fluctuations to assist in the creation of exciton condensates \cite{Johansen:PRL2019}.

We will review the recent developments in spin insulatronics. While magnonics  \cite{7kruglyak}, the exploration of spin waves in magnetic structures, is part of this emerging field  \cite{8cumak}, we focus on the new developments and possibilities exlusively in insulators enabled by electrical and thermal drives and detection in neighboring metals  \cite{Kajiwara:nat2010}. These new paths sometimes involve magnons in conventional ways as in magnonics and in new ways as mediators of attractions between electrons. Beyond magnons, there are also other collective ways to convey spin information in insulators. The title of our review, spin insulatronics, reflects the broader scope of these developments and that we focus on spins in insulating materials with ultra-low damping enabling new phenomena. We will consider the semiclassical regime of the generation, manipulation, and detection of low-energy coherent and incoherent spin waves and the collective quantum domain. Furthermore, super spin insulatronics constitutes magnon condensation and spin superfluidity, dissipationless transport of spin, which in (anti)ferromagnetic insulators may occur at room temperature. It also contains magnon-induced superconductivity, where the focus is on the new properties of the metals in contact with magnetic insulators.

\section{\label{sec:magneticinsulators}Magnetic Insulators}

Ferromagnets are often electrical conductors since ferromagnetic exchange relies on electron delocalization. Conversely, magnetic insulators are usually governed by indirect antiferromagnetic superexchange. The coupling generates either pure antiferromagnets or ferrimagnets when the magnetizations of the different sublattices do not entirely compensate each other. Tuning (or doping) of  the two (or more) sublattice structures and adjustment of their magnetization from zero for antiferromagnets to rather large values on the order of 100 kA/m are sometimes possible. In magnetic insulators, the only way to carry a spin current is via the localized magnetic moments. The spin flow is the propagation of their local disturbance. In its simplest manifestation, the spin current propagates via spin-waves (SWs), or their quanta magnons. The characteristic frequencies range from GHz to THz and the associated wavelengths range from $\mu$m to nm \cite{Gurevich:CRC96}.  A key feature of magnetic materials is that the SW dispersion relation can be continuously tuned by an external magnetic field over a very wide range. Additionally, changing or controlling the material alters the magnetic anisotropy, which provides additional means for tuning the spin transport properties.

In ferrimagnets, high-frequency excitations (up to THz) exist naturally, not only at very short wavelengths, through excitation of the magnetic moments in the two sublattices. Antiferromagnets are ferrimagnets with no net magnetization and an associated absence of the low-energy dispersion branch. While all magnetic insulators are useful in spin insulatronics, AFIs are of particular high interest. Their THz response is a real advantage and can facilitate ultrafast spintronics devices. They are also robust against an external magnetic field. Early theories of antiferromagnetic metals as active spintronic elements  \cite{11nunez,12haney} inspired their validation as memory devices \cite{13wadley,14marrows}. Significantly for our focus, AFIs have been demonstrated to be good spin conductors \cite{3lebrun,15hahn,16wang,Hou:NPGAsia2019}. The first pieces of evidence were indirect. Relatively short-range spin current propagation occurs in AFIs coupled to FIs \cite{15hahn,16wang}. The spin transport properties can vary by orders of magnitude by external magnetic fields\cite{Qiu:NatMat2018}. Importantly, truly long-range spin transport in an AFI, hematite, has been reported more recently \cite{3lebrun}. AFIs can also be good sources of spin currents \cite{li:nature2020,vaidya:science2020}. Other possible materials for room-temperature operation are the archetype NiO, CoO, and the magnetoelectric materials Cr\textsubscript{2}O\textsubscript{3} and BiFeO\textsubscript{3}.

Many insulating materials can function as spintronic elements. However, this variety is predominantly unexplored, as most published reports utilized yttrium iron garnet Y\textsubscript{3}Fe\textsubscript{5}O\textsubscript{12} (YIG). YIG is a ferrimagnetic insulator with the lowest known amount of spin dissipation as characterized by its exceptionally small Gilbert damping constant. Therefore, this material is optimal for propagating SWs. At low energies, the excitations in YIG resemble that of a ferromagnet. In ferromagnets, the excitations frequencies are set by the magnetic anisotropy that is typically much smaller than the exchange energy. The ferromagnetic resonance energies are therefore in the GHz regime in contrast to the THz excitations in antiferromagnets. The well-mastered growth of YIG thin films, either by liquid phase epitaxy \cite{17hahn,18dubs,19beaulieu}, pulsed laser deposition \cite{20sun,21kelly,22onbasli,hauser:scirep2016}, or sputtering \cite{wang:prb2013,chang:maglett2014}, so far has prevented other materials from being competitive. Surprisingly, spin propagation across microns in paramagnetic insulators with a larger spin conductivity than YIG has recently been reported \cite{23oyangi}. These results challenge the conventional view of spin transport, and we encourage a further elucidation of its fundamental origin.

Insulating ferrimagnets also potentially represent a fantastic playground for domain wall dynamics and the design of devices derived from the racetrack memory. Indeed, we can harvest two crucial exchange interactions to optimize the dynamical properties. Firstly, the antiferromagnetic (super-) exchange can play a role similar to its influence in antiferromagnets. Secondly, it is possible to induce a Dzyaloshinskii-Moriya interaction (DMI) through a well-chosen interface, mainly using the properties of direct contact with heavy metal \cite{Avci:NatNan2019,Velez:NatCom2019,Lee:PRL2020}. Rotating the local moments from their equilibrium position, generates several torques \cite{Ivanov:LTPhys2019} including a longitudinal field torque derived from the DMI field, a damping torque proportional to the time derivative of the moments, and that coming from the anisotropy field. For ferrimagnets, there is an additional torque derived from the exchange coupling field. All these torques contribute to the domain wall (DW) motion, but their amplitudes can be very different. Generally, damping and anisotropy field torques are small. The DMI torque can be larger in systems with well-chosen interfaces, but the exchange coupling torque is by far the greatest. It thus provides the main driving mechanism for the DW. The DW dynamics can be modeled by rescaling the gyromagnetic ratio and damping constant \cite{Avci:NatNan2019}. This approach results in an equation of motion for the Néel vector describing the collective dynamics. Unlike ferromagnets, DWs move very fast because saturation takes place at much higher current densities. This high current-driven mobility enabled by antiferromagnetic spin dynamics allows for DW velocities reaching 800 m/s for a current density around $1.2 \times 10^{12}$ A/m$^2$ \cite{Avci:NatNan2019}. These exciting properties are achieved (again) in garnets, particularly Tm$_3$Fe$_5$O$_{12}$ (TmIG), when in a bilayer with Pt. They are intimately linked to the interface-induced topology of the DWs. The presence of interfacial Dzyaloshinskii–Moriya interaction in magnetic garnets offers the handle to stabilize chiral spin textures in centrosymmetric magnetic insulators. Interestingly, the domain walls of TmIG thin films grown on Gd$_3$Sc$_2$Ga$_3$O$_{12}$ exhibit left-handed Néel chirality, changing to an intermediate Néel–Bloch configuration upon Pt deposition \cite{Velez:NatCom2019}. The DMI interaction seems, therefore, to emerge from both interfaces, but their exact balance is still unclear \cite{Avci:NatNan2019,Velez:NatCom2019,Xia:APL2020}. Still, recent measurements of topological Hall effect in ultra-thin TmIG/metal systems reveal the crucial role played by the FI/heavy metal interface \cite{Lee:PRL2020}.

Concepts related to the topology in condensed matter physics have been developed intensely in the past fifteen years. In magnetic materials, topology often arises as a consequence of spin-orbit coupling (SOC) and the breaking of inversion symmetry as typically exemplified by the Dzyaloshinskii-Moriya interaction. When this interaction is strong, it is sometimes able to curl the spins into hedgehog-type spin arrangements carrying a finite topological charge, the magnetic skyrmions. Their practical implementation in the field of spintronics is an important part of a new research area called spin-orbitronics. Magnetic skyrmions are emerging as a potential information carrier \cite{Bogdanov:JETP1989,Roessler:Nature2006}. Importantly, we can manipulate these topologically robust nanoscale spin textures with low current density \cite{Schultz:NatPhys2012,Nagaosa:NatNano:2013,Woo:NatMat2016,Nayak:Nature2017,Caretta:NatNan2018}. So far, ferromagnetic skyrmions are the only, most reliable and stable, topological quasi-particles (solitons) at room temperature in real-space condensed matter. They were originally discovered at low temperature and under a large field in bulk metallic non-centrosymmetric compounds \cite{Muhlbauer:Science2009},  but the community rapidly realized that we could generate them in metallic multilayers with broken inversion symmetry \cite{Heinze:NatPhys2011,Moreau-Luchaire:NatNano:2016,Boulle:NatNano:2016}.  These skyrmion systems, based on thin films or multilayers, offer more flexibility, room-temperature operation in the hope of being used as ultrasmall bits of magnetic information for mass data storage. As a first step towards a prototypal skyrmion based device, current-induced displacement was demonstrated \cite{Fert:NatNano2013,Jiang:NatPhys2017}. However, these metallic systems present several drawbacks, including too large power consumption. Therefore, it is technologically appealing for low-power information processing to avail of skyrmions in insulators, due to their low damping and the absence of Ohmic loss. It is also of fundamental interest for studying magnon-skyrmion interactions \cite{Schutte:PRB2014}. There are observations of skyrmions in the bulk of one insulating material: multiferroic Cu$_2$OSeO$_3$, albeit at low temperatures \cite{Seki:Science2012}. There are very recent reports that skyrmions can be generated at room-temperature in TmIG/Pt bilayers \cite{Shao:NatEle2019}. This very new avenue for insulating skyrmion-based room temperature systems will certainly offer opportunities for the next-generation of energy-efficient and high-density spintronic devices.

\section{Spin Injection and Detection in Magnetic Insulators}

\begin{figure}[htbp]
\includegraphics[width=0.99\columnwidth]{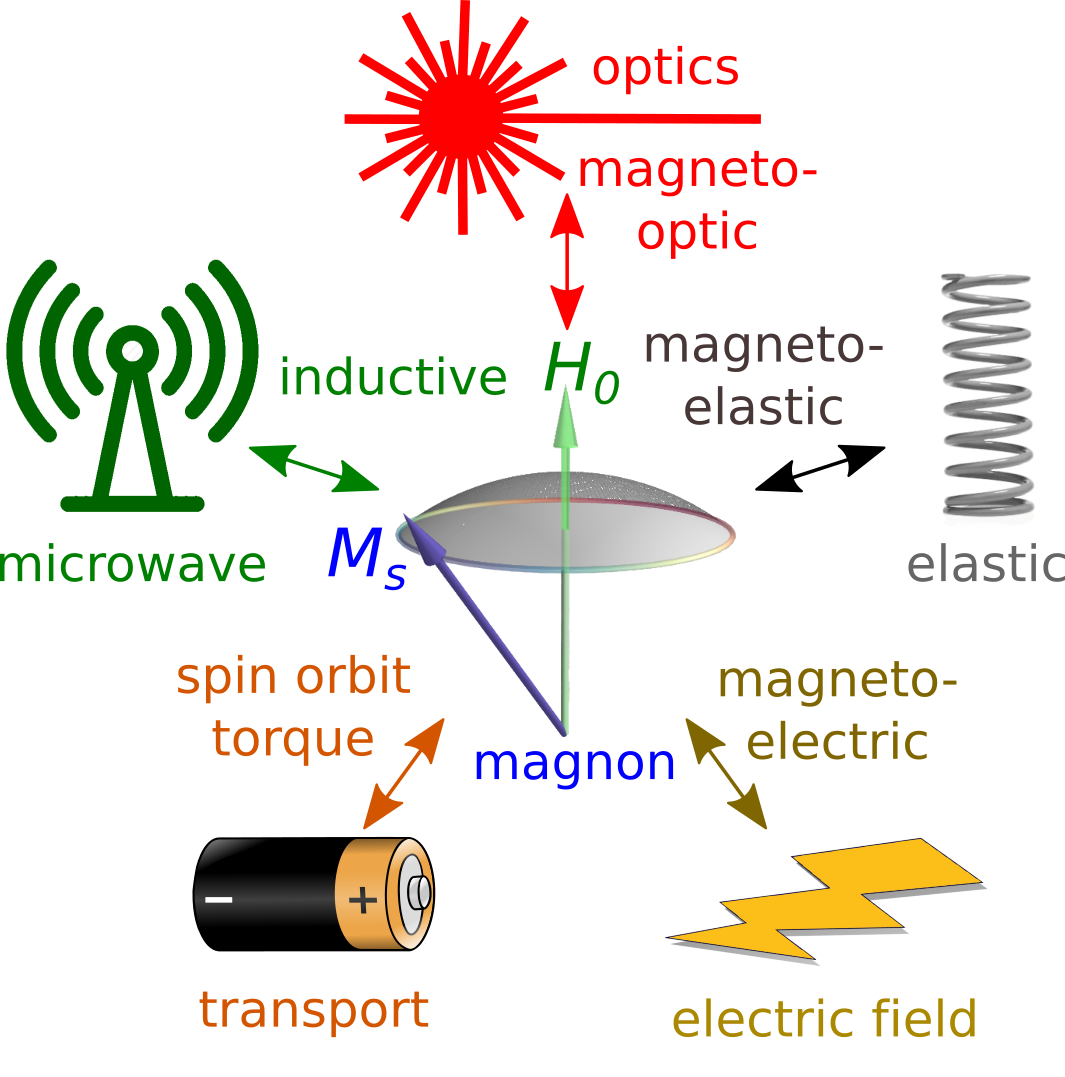}
%\vspace{-8pt}
\caption{Different ways to control the spin in magnetic materials. Inductive coupling with microwaves, elastic vibrations, optical irradiations, and electric fields causing spin-orbit torques exert torques on the magnetic moments.}
%\vspace{-8pt}
\label{fig:control}
\end{figure}

Insulators, in contrast to their metallic counterparts, have no (polarized) conduction electrons that can inject, detect or transport spin angular momentum. The absence of this simple link between charge and spin requires other types of couplings to the magnetic system. Spin insulatronics aims to deliver, control and eventually measure electric signals associated with spins in insulators. Fig.\ \ref{fig:control} provides a summary of the five interconversion processes that one could use in insulators to interconvert the spin information and another signal type: inductive, magnetooptical, magnetoelastic, magnetoelectric, and spin-transfer processes. All of these processes have been envisioned, but thus far, in spin insulatronics, most reported measurements have used electric current-driven spin current injection and detection. Before focusing on this approach, we present alternative means, including strain, light, electric and magnetic fields.

%%%%%%%%%%%%%%%%%%%%%%%%%%%
%%%%%%%%%%%%%%%%%%%%%%%%%%%%

\subsection{Probing Spins via Strain, Light, Electric and Magnetic fields}

The classical way to excite and detect SWs is inductive coupling. To excite SWs, a wire carrying an electric current is placed next to the magnetic insulators. In turn, the current induces a magnetic field via Ampere`s law. Similarly, to detect SWs, we can measure electric currents induced by the stray fields of the SWs. However, the inductive coupling approach has shortcomings when applied to the microscopic scale exploited by spin. The spatial resolution of the nanopatterning technology exploited to fabricate a microwave antenna limits the smallest wavelength that can be excited and detected \cite{24yu}. In practice, the resolution limit is in the submicron range. Therefore, the inductive scheme only addresses SWs with energies that are orders of magnitude below those of thermal magnons. More importantly, the coupling is a volumetric effect. The wavelength mismatch between the microwaves and SWs makes the process inherently inefficient in detecting SWs in thin films \cite{25collet}, implying either poor sensitivity or an inability to induce large power excitation.

Applying strain is a method for controlling magnetic excitations. Magnetostriction is very common in magnetic materials \cite{26tremolet}. In antiferromagnets, strain can also couple to the Néel vector through linear magnetoelastic coupling \cite{27slack,28greenwald}. The ability to trigger collective excitations using magnetoelastic coupling \cite{29dreher} and to interconvert these excitations and elastic waves \cite{30kittel} through interdigitated piezo-electric transducers is an alternative way to detect angular momentum \cite{31holanda,32zhang}. Spin-pumping in YIG using acoustic waves providing an electrically tunable source of spin current \cite{33matthews}, albeit at rather low frequencies (MHz), has also been demonstrated. The use of magnon polarons processes \cite{34kikkawa} enables investigations of high-frequency spin transport in ferromagnetic materials. Moreover, the development of high-frequency coherent shear acoustic waves \cite{35litvinenko} has opened opportunities for coupling magnons in antiferromagnets with acoustic phonons \cite{36watanbe,37ozhogin}. 

The issue of weak coupling to the magnetic information is the drawback of magneto-optical techniques \cite{38tabuchi,39jorzick}. Nevertheless, hybridization of whispering gallery optical modes with Walker SW modes propagating in the equatorial plane of a YIG sphere was recently demonstrated \cite{40osada}. The development of ultrafast light sources has enabled triggering and detection of magnetic excitations in ferromagnets and antiferromagnets in the time domain.  This control is achieved using ultrafast femtosecond lasers \cite{41reid} in a pump-probe fashion, through different interactions. Ultrafast shock wave generation \cite{42kim,43ruello} can generate magnons, either directly inside the material of interest or with the help of a thin layer of a transducer material \cite{44kovalenko}. The injection of spin currents through ultrafast demagnetization \cite{45choi} or the even more direct inverse Faraday effect \cite{46vander,47mikhaylovskiy,48satoh} is also an important phenomenon. Another photoinduced  mechanism is the ultrafast change of anisotropy used for triggering SWs in NiO \cite{49duong,50rubano}, as well as the direct torque induced by the magnetic component of THz monocycle pulses \cite{51kampfrath}. Noticeably, many of the pump-probe studies have been performed on bulk insulators because optimal dynamical properties require the low damping obtained when electrical currents cannot flow. The detection of signals from the magnetic excitations often requires a significant sample volume. Thus, measurements carried out on thin films in the framework of spintronics and magnonics remain scarce. Moreover, precise control of the SW emission using these techniques is currently lacking.

In materials where magnetic and electric orders are intrinsically coupled, pure electric fields can enable electric control of magnetic properties. Such magnetoelectric phenomena have been at the center of research since the turn of the century \cite{52eerendtein}. These 'multiferroic' compounds have rich physics and are potentially appealing for applications, especially when they are ferromagnetic and ferroelectric \cite{53hill}. Significant magnetoelectric coupling between the two order parameters allows for magnetic manipulation of the ferroelectricity or, conversely, electrical control of the magnetic order parameter \cite{54fiebig,55hur}. A common way of expressing this coupling is by introducing terms into the free energy coupling of the polarization and the magnetization (or, more often, the antiferromagnetic Néel vector). The coupling can be linear or quadratic \cite{56pyatakov}. Most magnetic multiferroics are antiferromagnets, including the archetypical BiFeO3 (BFO), the only multiferroic with its two ordering temperatures well above 300K \cite{57catalan}. Rotating the ferroelectric polarization in this compound results in a rotation of the sublattice Néel vector \cite{58zhao,59lebeungle}, a property used to design low-consumption memories \cite{60allibe}. Surprisingly, the dynamical properties of the magnetoelectric coupling and its utilization in the design of magnonic structures are underexplored. Conceptually, it has been shown that magnons in magnetoelectrics can be hybrid entities because of their coupling to the electric order \cite{61rovillain,62pimenov}. In principle, one can envision that electric fields can launch the resulting 'electromagnons' to generate and control magnonic transport at the micron scale, a feature not yet demonstrated. In any case, the magnetoelectric effect could be very useful in the field of spin insulatronics, particularly in addressing and controlling antiferromagnets, for which magnetic fields are inoperable.

\subsection{Magneto-elastic Coupling}

Interest in magneto-elastic properties at high frequencies started at the end of the 50s. The landmark is probably the seminal paper of Kittel \cite{30kittel}, who explained the benefits of coupling acoustic-waves (AW) with spin-waves (SW) in order to confer to the former waveform a tunable and non-reciprocal character. Since then, the subject has accumulated a vast amount of literature that spans a wide range of ferromagnetic materials from metals \cite{Seavey:IEEE1963}, to magnetic semiconductors \cite{Thevenard:PRB2014,Thevenard:PRB2016}, and electrical insulators \cite{Fine:PR1954,Gibbons:JAP1957,Spencer:PRL1958}. From the start, ferrite garnets \cite{Lecraw:1965} have been the subject of an intense focus because of their ultra-low internal frictions. These features are beneficial not only to the magnetization dynamics but also to the ultrasonic propagation. Benchmarking the latter performance, it turns out that the sound attenuation in garnets (both YIG and GGG) is ten times lower than in quartz \cite{Lecraw:1965}. This remarkable property led to an intense research program in the 60s, in particular at Bell Labs. Spencer and LeCraw envisioned the interest of developing tunable delay lines relying on gyromagnetic effect as front-end microwave analog filters for heterodyne detection of wireless signals. Spencer and LeCraw's efforts were thorough. They went as far as putting a YIG sphere in magnetic levitation \cite{Lecraw:1965} in order to minimize contact loss of sound waves to investigate the limits of acoustic decays in these materials. Eventually, garnets delay lines never reached market products principally because of their reduced piezoelectric coefficient, which implies substantial insertion loss for phonon interconversion, which are orders of magnitude higher than their non-magnetic counterpart.

The subject of phononic currents inside magnetic insulators gained a renewed interest with the recent developments of spintronics \cite{Streib:PRL2018,an:PRB2020} and in quantum information technology. Current research focuses on two properties of the magneto-elastic coupling. The first is the ability to reach the strong coupling regime between magnons and phonons, which implies the coherent transfer of information \cite{Bienfait:Science2019}. The second is the ability of circularly polarized phonons to carry angular momentum \cite{30kittel,Garanin:PRB2015,Holanda:NatPhys2018,rueckriegel:PRB2020,rueckriegel:PRL2020}, allowing sound waves to carry spin information across vast distances that exceed previous benchmarks set by magnon diffusion \cite{2cornelissen,3lebrun} by orders of magnitude.

\begin{figure}[htbp]
\includegraphics[width=0.99\columnwidth]{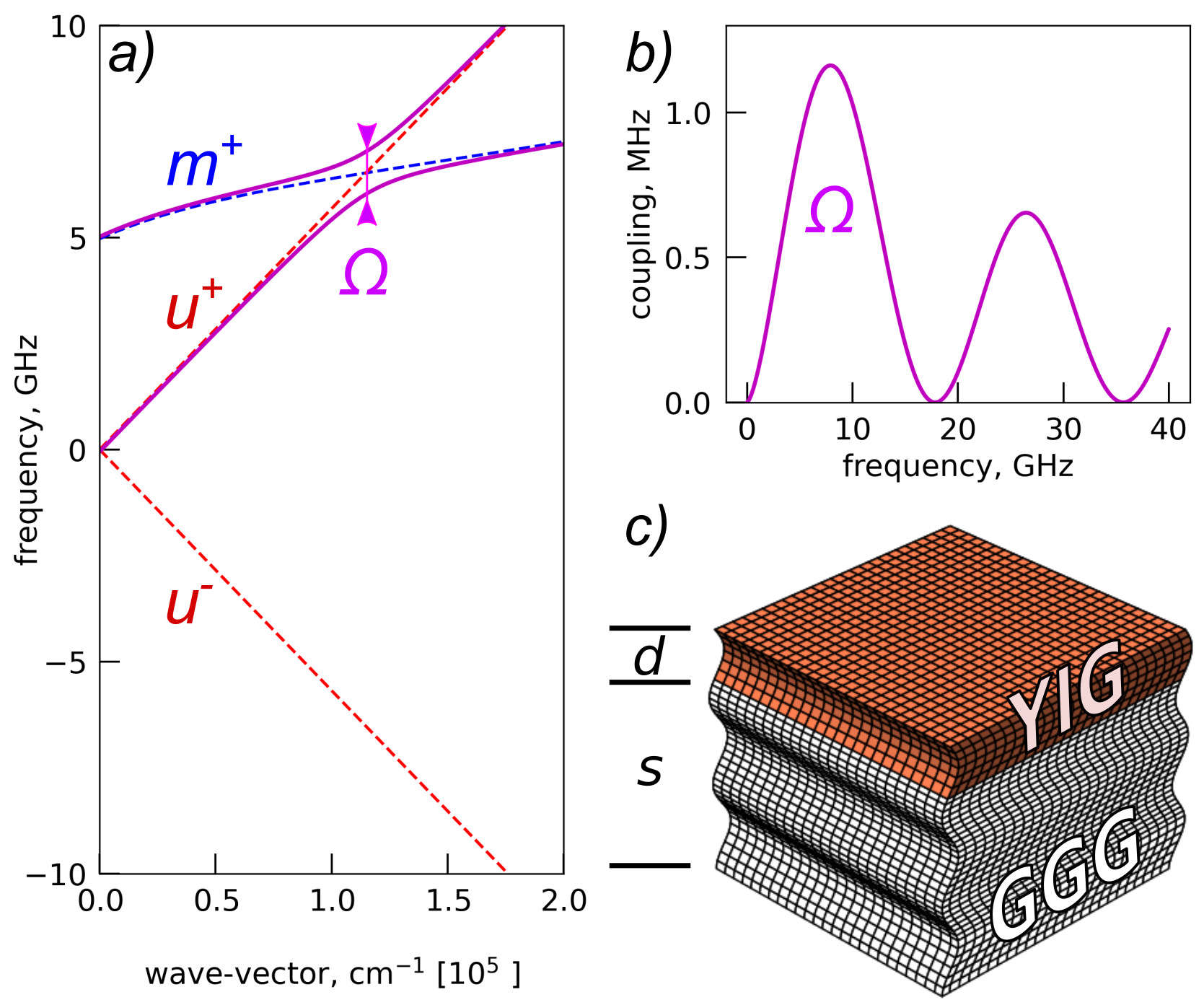}
%\vspace{-8pt}
\caption{ 
a) Schematic representation of the dispersion relation of acoustic waves (AW) and spin-waves (SW). Magneto-elasticity leads to new hybrid quasi-particles, "magnon polarons", when spin-wave  and acoustic-wave dispersions anti-cross. Note that a linearly polarized field can always be considered as a superposition of two circularly polarized fields $u^+$ and $u^-$ rotating in the opposite direction. Since spin-waves are circularly polarized vectors rotating around their equilibrium direction anti-clockwise (or in the gyromagnetic direction), the coherent coupling confers to the acoustic waves a finite angular momentum. b) Frequency dependence of the overlap integral $\Omega$ defined by Eq.\ (\ref{eq:Omega}). c) Schematic illustration of coherent shear waves propagating in a perpendicularly magnetized thin film of thickness $d$ on a substrate of thickness $s$.
}
%\vspace{-8pt}
\label{fig:magnonphononhybrid}
\end{figure}

In ferromagnets, the hybridization between spin-waves (magnons) and lattice vibrations (phonons) stems from the magnetic anisotropy and strain dependence of the magnetocrystalline energy \cite{30kittel,Bommel:PRL1959,Damon:IEEE1965,Seavey:IEEE65}. It implies that crystal growth orientation and magnetic configuration determine the selection rules for the coupled eigenvectors. In the following, we shall concentrate on thin films magnetized along the normal direction ($z$-axis). In this case, the dominant coupling with the Kittel mode is achieved by coherent acoustic shear waves. The coupled equation of motion, 
\begin{equation}
(1-i \alpha_m ) \omega m^+ = \gamma (H m^+ - D \partial_z^2 m^+ + B \partial_z u^+ - M_s h^+)
\label{eq:m+}
\end{equation}
and
\begin{equation}
    -(1-2 i \alpha_a) \omega^2 \rho u^+ = C_{44} \partial_z^2 u^+ + \frac{B}{M_s} \partial_z m^+
    \label{eq:u+}
\end{equation}
are governed by the vertical derivatives of the rotating fields. In Eqs.\ (\ref{eq:m+}) and (\ref{eq:u+}), $m^+$, $u^+$,  and $h^+$ are respectively the circularly polarized oscillating part of the magnetization, lattice displacements, and rf magnetic field.  $H$ is the effective bias magnetic field, comprising the externally applied field, the anisotropy and the demagnetizing magnetic field; $D$ is the exchange constant; $B=(B_2+2B_1 )/3=7 \times 10^5$~J/m$^3$ is the effective coupling constant, with $B_1$ and $B_2$  being the magneto-elastic coupling constants for a cubic crystal; $\rho =5.1$~g/cm$^3$ is the mass density; $\alpha_m$ and $\alpha_a$ are respectively the magnetic and acoustic damping coefficient; and finally $C_{44}$ is the elastic constant, which governs the transverse sound velocity through the relation $v=\sqrt{C_{44}/\rho}$ (see chapter 12 of Ref.\ \cite{Gurevich:CRC96}). The set of coupled equations (\ref{eq:m+}) and (\ref{eq:u+}) is more easily expressed by linearizing Eq.\ (\ref{eq:u+}) around the Kittel frequency $\omega_m $ and integrating the coupling over the total crystal thickness. The integration will be done for the case of a phononic bi-layer system of a YIG thin film grown on top of a GGG substrate, a non-magnetic dielectric (see Fig.\ \ref{fig:magnonphononhybrid}c). We will neglect any impedance mismatch between the two layers, and for any practical purpose the sound properties are governed by the total crystal thickness. In our notation, $d$ and $s$ are respectively the thickness of the magnetic layer (indicated in red) and the substrate (indicated in grey) in Fig.\ \ref{fig:magnonphononhybrid}c. Assuming that both waveform $m^+$ and $u^+$  are plane waves, the coupled equations of motion simplify to the standard form
\begin{equation}
    (\omega - \omega_m + i \alpha_m \omega) m^+ = \frac{\Omega}{2} u^+ + \kappa h^+
    \label{eq:msimple}
\end{equation}
and
\begin{equation}
(\omega - \omega_a + i \alpha_a \omega) u^+ = \frac{\Omega}{2} m^+
\label{eq:usimple}
\end{equation}
where $\kappa$ is the inductive coupling to a microwave antenna and $\Omega$ is the magneto-elastic overlap integral between the AW and the SW
\begin{equation}
    \Omega = \frac{B}{\sqrt{2}} \sqrt{\frac{\gamma}{\omega M_s \rho s d}} \left(1 - \cos{\omega \frac{d}{v}} \right) \, .
    \label{eq:Omega}
\end{equation}

%%%%%%%%%%%%%%%%%%%%%%%%%%%
%%%%%%%%%%%%%%%%%%%%%%%%%%%%

 Although the absolute value of  $B$ is usually quite small inside garnets, the strong coupling regime is possible because $B$ is comparable to the damping rates of two hybrid waveforms. To illustrate this point, we have traced in Fig.\ \ref{fig:magnonphononhybrid}b the frequency variation of the overlap integral $\Omega$ for a $d=200$~nm thick YIG film deposited on top of a $s=0.5$~mm GGG substrate. The oscillating behavior of $\Omega$ translates the fact that the optimal coupling occurs when the film thickness reaches the half-wave condition. Consequently, there are sweet spots in the frequency spectrum at which the coupling $\Omega$ is maximal, and these occur at finite rf frequencies ($\Omega=0$ at $\omega=0$). In the case of our 200 nm thick YIG film, the first maximum occurs at about 6 GHz.

Reaching the strong coupling regime requires that the dimensionless quantity formed by the ratio of the coupling strength squared to the product of the relaxation rates of each hybrid waveform, $C= \Omega^2/(2 \eta_a \eta_m)$, called the cooperativity, becomes greater than 1. The associated signature of reaching the strong coupling regime is the avoided crossing of energy levels, as shown in Fig.\ \ref{fig:magnonphononhybrid}a, which proves that a coherent hybridization between the two waveforms occurs leading to the formation of new hybrid quasiparticles called ``magnon polarons''. Since, in our case, the excitation of the magnetic order conserves the axial symmetry, the magnetic excitation must be purely circularly polarized, as shown in Fig.\ \ref{fig:magnonphononhybrid}a. The magneto-elastic interaction (a conservative linear tensor) transfers the polarization faithfully to the elastic shear-wave deformation. Thus the circular polarization of the phonons is a consequence of the axially symmetric configuration and the associated conservation of angular momentum (Noether theorem).

\begin{figure}[htbp]
\includegraphics[width=0.99\columnwidth]{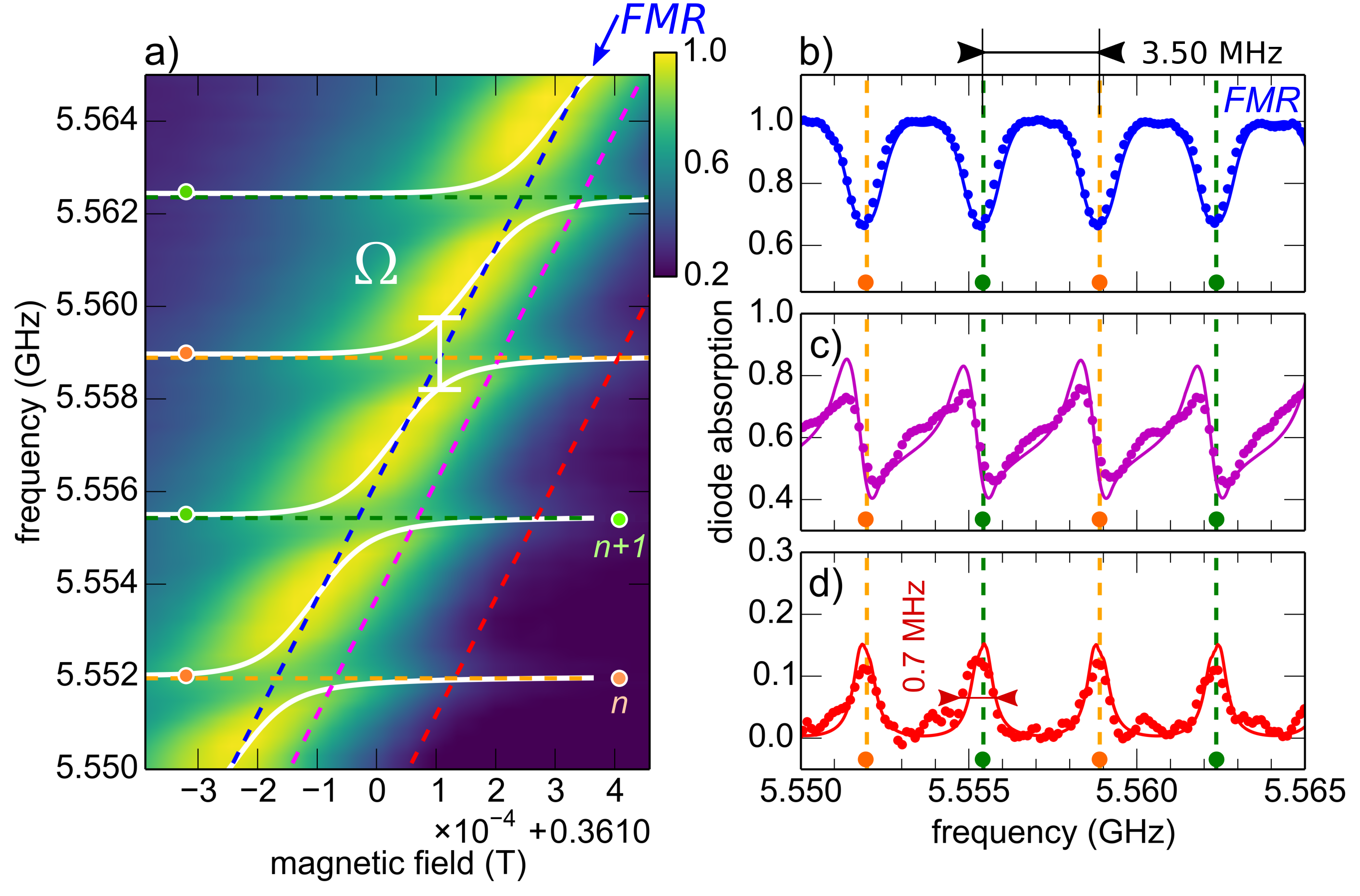}
%\vspace{-8pt}
\caption{ 
a) Density plot of the ferromagnetic absorption signal of a YIG(200 nm)|GGG(0.5 mm) epitaxial crystal, revealing the anti-crossing between the FMR mode (indicated by blue arrow) and the coherent standing shear acoustic waves that resonates across the total crystal thickness. These acoustic resonances form a regular comb of absorption lines at $\omega_n=n \pi v/(d+s)$, where n is their mode number, thus producing a periodic modulation at 3.50~MHz represented in the figure as horizontal dash lines in orange and green. The right panels (b,c,d) show the intensity modulation for 3 different cuts (blue, magenta and red) along the gyromagnetic ratio, i.e., parallel  to the resonance condition. The solid lines in the 4 panels are fits by the coupled oscillator model (cf. Eqs.\ (\ref{eq:msimple}) and (\ref{eq:usimple})). Figure from ref.\cite{an:PRB2020}.
}
%\vspace{-8pt}
\label{fig:densityplotFMR}
\end{figure}

In the following, we shall illustrate experimentally that indeed it is possible to reach the strong coupling between the Kittel mode and coherent acoustic shear waves by working at high frequency in the normal configuration  \cite{an:PRB2020}. The crystal is an epitaxial film of YIG of 200nm thickness deposited on top of a 0.5 mm GGG substrate. Fig.\ \ref{fig:densityplotFMR} shows the FMR absorption of the Kittel mode around 5.56 GHz. These spectra are in the perpendicular configuration, where the magnetic precession is precisely circular. A fine periodic structure appears in the absorption spectrum when one performs a very fine scan of both the external magnetic field and the frequency. A canted arrow indicates the Kittel relationship linking the field and the frequency labeled FMR. The 3.50 MHz modulation periodicity in the absorption signal due to standing phonon modes that resonate across the whole crystal of thickness $s+d$. The quantification of their wavevector is governed by the transverse sound velocity inside GGG along (111) of $v=3.53 \times 10^3$~m/s \cite{Ye:PRB1991}  via $v/(2d+2s) \approx 3.53$~MHz. At 5.5 GHz, the intercept between the transverse AW and SW dispersion relations occurs at $2 \pi/ \lambda n = \omega/v \approx 10^5$ cm$^{-1}$, which corresponds to a phonon wavelength of about $\lambda_n \approx700$~nm with index number $n \approx 1400$. The modulation is strong evidence for the high acoustic quality that allows elastic waves to propagate coherently with a decay length exceeding twice the substrate thickness, which is 1 mm in this case.

%%%%%%%%%%%%%%%%%%%%%%%%%%%
%%%%%%%%%%%%%%%%%%%%%%%%%%%%

We also show in Fig.\ \ref{fig:densityplotFMR} what happens to the line shapes when one records the modulation at detunings parallel to the FMR resonance as a function of field and frequency indicated by the blue, magenta, and red cuts. The amplitude of the main resonance (blue) in Fig.\ \ref{fig:densityplotFMR}b dips and the lines broaden at the phonon frequencies \cite{Ye:PRB1991}. The minima transform via dispersive-looking signal (magenta) into peaks (red) once sufficiently far from the Kittel resonance as expected from the complex impedance of two detuned resonant circuits, illustrating a constant phase between $m^+$ and $u^+$  along these cuts. Such a behavior is already the signature of a coherent coupling between the two waveforms.

The observed line shapes can be used to extract the lifetime parameters of both the acoustic and magnetic waveform. The 0.7 MHz full line width of the acoustic resonances in Figure 2d indicates the
AW lifetime $\eta_a/(2 \pi) =0.35$~MHz as the half line width of the acoustic resonance. The SW lifetime $1/\eta_m$  follows from the broadening of the absorbed power at the Kittel condition which contains a constant inhomogeneous contribution and a frequency-dependent viscous damping term $\eta_m= \alpha_m \omega_m$. When plotted as function of frequency, the slope gives the Gilbert phenomenology of the homogeneous broadening  corresponds to a $\eta_m/(2 \pi) = 0.50$~MHz at 5.5 GHz.

A remarkable feature in Fig.\ \ref{fig:densityplotFMR}a are the clearly resolved avoided crossings of SW and AW dispersion relations, which prove the strong coupling between two oscillators. Fitting by hand the dispersions of two coupled oscillators through the data points (white lines), we extract a gap of $\Omega /(2 \pi)=1$~MHz and thus a large cooperativity $C \approx 3$  in agreement with the observation of avoided crossing. This coupling value can then be injected in the set of coupled equations (\ref{eq:msimple}) and (\ref{eq:usimple}) to infer the expected behavior at the various detunings. The results are shown as solid lines in the three panels of Fig.\ \ref{fig:densityplotFMR} bcd. Such an agreement between the data and the model establishes that efficient power transfer can be achieved between the magnon tank and the phonon tank.

These findings unambiguously show that magnets can act as source and detector for phononic angular momentum currents and they suggest that these currents should probably be able to provide a coupling, analogous to the dynamic coupling in metallic spin valves, but with an insulating spacer, over much larger distances, and in the ballistic/coherent rather than diffuse/dissipative regime \cite{an:PRB2020,rueckriegel:PRL2020}.  These findings might also have implications on the non-local spin transport experiments\cite{2cornelissen,116cornelissen,139cornelissen}, in which phonons provide a parallel channel for the transport of angular momentum.

\subsection{Spin Hall Effect}

The recent discovery of spin-transfer and spin-orbit effects enables injection of an external spin current through the interface from an adjacent nonmagnetic layer \cite{63brataas}. This method provides direct electric control of spin transport and has overcome many limitations of previously established routes. We will, therefore, focus on this method. Electric currents passing through conductors can generate pure spin signals. In metals with a significant spin-orbit interaction, such as Pt and W, the spin Hall effect (SHE) converts a charge current into a transverse spin current \cite{64sinova}. The generated spin current is 
\begin{equation}
    j_{\alpha \beta}^{(s)}=\theta_{s H} \varepsilon_{\alpha \beta \gamma} j_{\gamma}^{(c)}
\end{equation}
where  \(  \theta _{sH} \)  is the spin Hall angle, and  \(  \varepsilon _{\alpha \beta \gamma} \)  is the Levi-Civita tensor. The charge current  \( j_{\gamma}^{ \left( c \right) } \)  is the component that flows along the  \( \gamma \)  direction. The spin current  \( j_{\alpha \beta}^{ \left( s \right) } \)  flows along the  \( \beta \)  direction and is polarized along the  \( \alpha \)  direction. We can in turn inject the generated transverse spin current into magnetic insulators. Detection of spin currents is also feasible. An effect reciprocal to the spin Hall effect exists, in which a spin current can cause a secondary transverse charge current via the inverse spin Hall effect (ISHE). In this, so-called inverse spin Hall effect, the generated charge current is proportional to the primary spin current
\begin{equation}
j_{\alpha}^{(c)}=\theta_{s H} \varepsilon_{\alpha \beta \gamma} j_{\beta \gamma}^{(s)} \, ,
\end{equation}
The magnitude of the spin Hall angle $\theta_{s H}$ is an important factor. Its deduced value differs in various experiments. Care should be taken in extracting values of the spin Hall angle from different experiments \cite{Nguyen:PRL2016,Zhu:PRL2019}.

Rasbha-coupled interfaces \cite{65sanchez} or topological insulators, such as Bi\textsubscript{2}Se\textsubscript{3}, also facilitate an analogous spin-charge coupling \cite{66mellnik}. The spin Hall magnetoresistance, the dependence of the resistance in metals on the magnetic configuration of adjacent insulators, can be used to probe ferromagnetic \cite{17hahn,67nakayama} and antiferromagnetic arrangements \cite{68aqeel}. 

\subsection{Spin-transfer, Spin-orbit torques, and Spin-pumping} 

Spin angular momentum can flow from metals to magnetic insulators or in the opposite direction via the exchange coupling at the metal-insulator interfaces. At these connections, the energy depends on the relative orientations of the localized and the itinerant spins. A disturbance in either of the spin subsystems can therefore propagate from metals into insulators and vice versa. In the metal, spin-polarized transport or spin-orbit coupling together with charge transport can cause a spin imbalance, resulting in a spin-transfer torque (STT) or spin-orbit torque (SOT), respectively. With the spin-transfer torques in ferromagnets, a spin accumulation, resulting, for instance, from the spin Hall effect, is transferred as a torque on the magnetization \cite{63brataas,89berger,90slonczewski,Waintal:PRB2000,Brataas:PRL2000,Brataas:EPJ2001,91stiles},
\begin{equation}
\tau=a m \times\left(m \times \mu^{(s)}\right)
\end{equation}
where  $a$  is a measure of the efficiency that is proportional to the transverse spin conductance (spin mixing conductance) \cite{Brataas:PRL2000,Brataas:EPJ2001},  $m$  is a unit vector along the magnetization direction, and  \(  \mu ^{ \left( s \right) } \)  is the spin accumulation. The reciprocal effect also exists. A dynamical magnet pumps spin-currents to adjacent conductors
\begin{equation}
j_{i z}^{(s)}=b m \times \frac{\partial m}{\partial t} \, ,
\end{equation}
where the spin-pumping efficiency $b$ is related to the spin-transfer efficiency $a$ \cite{81tserkovnyak,83tserkovnyak}. While spin-pumping and spin-transfer torque are reciprocal effects, in insulators, the former effect is easier to measure. Analogous expressions exist for the spin-transfer torques in AFIs \cite{69cheng} where the unit vector along the magnetization direction can be replaced by the unit vector along the Néel vector.

A broad range of experiments on a variety of systems have unambiguously established spin-pumping in insulating ferromagnets \cite{70sandweg,71sandweg,72vilela,73rezende,74azenvedo,75burrowes,76kurebayashi,77kurebayashi,78hahn,wang:prb2013}. Theoretically, spin-pumping is predicted to be as strong from FIs \cite{80kapelrud} and AFIs \cite{69cheng,Kamra:PRL2017,79johansen} as from ferromagnetic metals \cite{81tserkovnyak,83tserkovnyak,82brataas}. This essential mechanism should, therefore, be robust. Indeed, very recently, two independent groups provided direction demonstrations of spin-pumping from AFIs \cite{li:nature2020,vaidya:science2020}.

Spin-transfer and spin-orbit torques provide new avenues to alter the magnon energy distribution in insulators \cite{84xiao,85du,86demidov}. However, the generation of magnons in these ways lacks frequency selectivity, and can therefore lead to excitations in a broad frequency range \cite{87hamadeh,88demidov}. This lack of selectivity poses a challenge in identifying the SW modes that are propagating the spin information. 

Although the selection rules of spin-transfer effects seem insensitive to the SW spatial pattern, the spin-transfer efficiency increases with decreasing magnon frequency \cite{89berger,90slonczewski,91stiles}. The energy transfer relies on a stimulated emission process \cite{89berger,90slonczewski}. By favoring the SW eigen-modes with the most substantial fluctuations, spin-transfer preferentially targets the modes with the lowest damping rates, the lowest energy eigen-modes since the relaxation rate is proportional to the energy \cite{88demidov,92haidar}. The situation is different for SWs excited by thermal heating, where the excitations predominantly consist of thermal magnons, whose number overwhelmingly exceeds the number of lowest energy modes \cite{93xiao,94hoffman}.

Additional opportunities arise from the improved efficiency of the spin-transfer process. Injecting strong spin currents into small areas and bringing the magnetization dynamics into the deep nonlinear regime are possible. Insulating magnetic materials are particularly promising since they have exceptionally low damping. Indeed, the relevant quantity is the amount of external spin density injected relative to the linewidth. Using spin-transfer in insulators provides a unique opportunity to probe nonequilibrium states, where new collective behaviors are expected to emerge, such as Bose-Einstein (BEC) condensation at room temperature \cite{95bender,96bender}. Importantly, nonlinear processes are responsible for energy-dependent magnon-magnon interactions, which lead to threshold effects such as SW instabilities \cite{97suhl}. Such phenomena can be described as turbulence, which is most well-known as the mechanism behind the saturation of the Kittel mode or the rapid decay of coherent SWs into incoherent motions \cite{Lvov:94}. As a consequence, these effects alter the energy distribution of magnons inside the magnetic body. These processes should conserve both energy and angular momentum. Therefore, a critical parameter that controls magnon-magnon interactions is confinement. In fact, finite size effects lift the degeneracy between modes, limiting the existence of quasi-degenerate modes available at the first harmonics, which substantially increases the nonlinear threshold values. In closed geometries (e.g., nano-pillars), when confinement fully lifts the degeneracy between the SW modes, spin-transfer processes can generate large coherent GHz-frequency magnon dynamics. In this case, a single mode tends to dominate the dynamics beyond the critical spin current for damping compensation \cite{84xiao}:
\begin{equation}
  J_s^*=\frac{1}{\gamma} \left(\frac{\partial \omega}{\partial H}\right) \frac{\alpha \omega M_s t_\mathrm{FI}}{\gamma}\, ,
  \label{eq:Jsc}
\end{equation}
in which the single SW mode is precessing at $\omega$ in the field $H$, and $\alpha$, $M_s$ and $t_\mathrm{FI}$ are respectively the damping, magnetization and thickness of the FI layer. This selection enables control of the amplitude. The demonstrations of current-induced torques affecting the magnons in YIG \cite{87hamadeh,92haidar,99collet,100demidov} use this feature.

%%%%%%%%%%%%%%%%%%%%%%%%%%%
\begin{figure}[htbp]
\includegraphics[width=0.99\columnwidth]{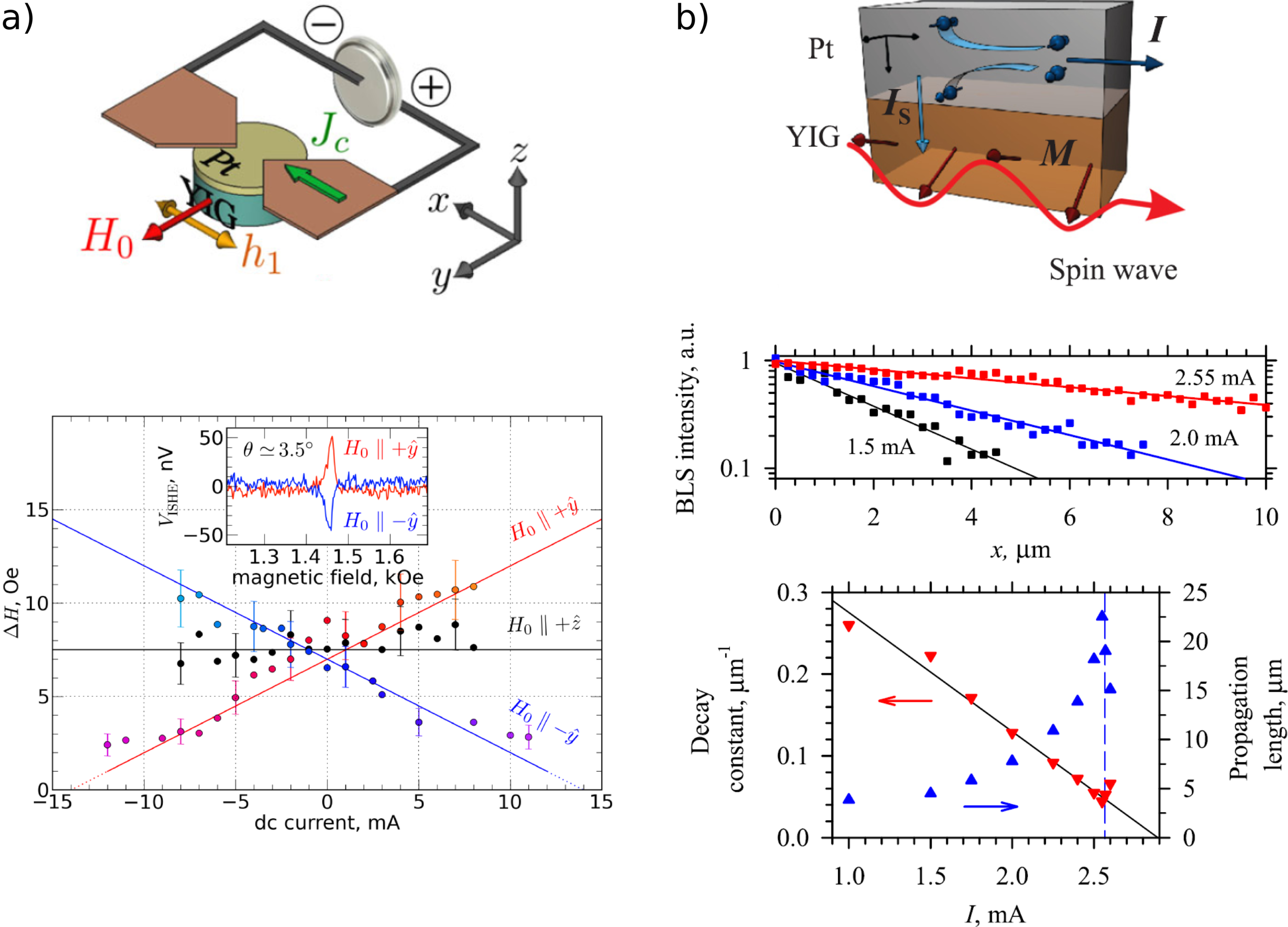}
%\vspace{-8pt}
\caption{ 
a) Electrical control of the magnetic damping in a YIG/Pt microdisk (from ref. \cite{87hamadeh}). Top: Schematics of the sample (a 5~$\mu$m diameter YIG(20~nm)/Pt(7~nm) disk) and measurement configuration. Bottom: Variation of the full linewidth measured at 6.33~GHz as a function of the dc current for out-of-plane field (black) and two opposite in-plane field (red and blue) configurations. The inset shows the detection of $V_\mathrm{ISHE}$ in the transverse direction of the in-plane swept magnetic field at $I_\mathrm{dc}=0$. b) Stimulated amplification of SWs by SOT (from ref. \cite{evelt:apl2016}). Top: Schematics of the sample (a 1~$\mu$m wide YIG(20~nm)/Pt(8~nm) stripe waveguide) and measurement configuration. Middle: SW decay for different currents in the Pt. Bottom: Current dependence of the decay constant and of the propagation length. A nearly 10 times increase of the latter is observed just before the auto-oscillation threshold, marked by the vertical line.
}
%\vspace{-8pt}
\label{fig:yigpt-damping}
\end{figure}
%%%%%%%%%%%%%%%%%%%%%%%%%%%%

To illustrate the effect of SOT from an adjacent heavy metal layer on the magnetization dynamics of FIs, we briefly discuss two seminal experiments on the control of magnetic damping in a YIG/Pt microdisk \cite{87hamadeh} (Fig.\ \ref{fig:yigpt-damping}a) and on the amplification of propagating SWs in a YIG/Pt waveguide \cite{evelt:apl2016} (Fig.\ \ref{fig:yigpt-damping}b). 

In the experiment depicted in Fig.\ \ref{fig:yigpt-damping}a, a YIG(20~nm)/Pt(7~nm) bilayer is patterned into a microdisk of diameter 5$~\mu$m to which gold electrodes are contacted for dc current injection. In order to maximize the effect of spin-orbit torque, the sample is biased in the plane, with the magnetic field $H_0$ oriented transversely to the dc current $J_\mathrm{c}$ flowing in the Pt layer. This is indeed the most favorable configuration to modulate the damping, as spins accumulated at the YIG/Pt interface due to SHE in Pt will be collinear to its magnetization (see also Fig.\ \ref{fig:sotseebeck}a and b).

The ferromagnetic resonance is excited in the YIG microdisk using the rf field $h_1$ produced by a microwave antenna patterned on top of the sample. The evolution of the in-plane (negative/positive bias) and out-of-plane standing spin-wave spectra of the YIG/Pt disk is monitored by a magnetic resonance force microscope as a function of the dc current injected into the 7~nm thick Pt layer \cite{87hamadeh}. The values of the full linewidth measured at 6.33~GHz in these geometries with respectively blue/red and black symbols, are reported in Fig.\ \ref{fig:yigpt-damping}a as a function of current. The blue/red data points follow approximately a straight line, whose slope $\pm0.5$~Oe/mA reverses with the direction of the bias field, and whose intercept with the abscissa axis occurs at $I^*=\mp14$~mA, which corresponds to a current density of about $4~10^{11}$A/m$^2$. The variation of linewidth covers about a factor five on the full range of current explored. In contrast, the linewidth measured in the out-of-plane geometry does not change with current. In fact, no net spin-transfer torque is exerted by the spin current on the precessing magnetization in this configuration.

The inset of Fig.\ \ref{fig:yigpt-damping}a displays the inverse spin Hall voltage $V_\mathrm{ISHE}$ measured at $I=0$~mA and $f=6.33$~GHz on the same sample, for the two opposite orientations of the in-plane bias field. This allows to check that a spin current can be generated by spin pumping and transmitted from YIG to Pt, and which polarity of the current is required to compensate the damping. In this experiment, damping compensation occurs when the product of $V_\mathrm{ISHE}$ and $I$ is negative. This is consistent with having a positive spin Hall angle in Pt.

The results of Fig.\ \ref{fig:yigpt-damping}a unambiguously demonstrate that SOT can be used to control the relaxation of a YIG/Pt hybrid device, and the current density extrapolated to reach zero linewidth is close to the one calculated using the expression of the critical current for damping compensation Eq.\ref{eq:Jsc} with independently determined experimental parameters. By solely biasing similar YIG/Pt microdisks with a dc current (no microwave excitation applied), auto-oscillations of the YIG magnetization were also observed beyond the critical current \cite{99collet}. We remind here the important role of finite size effects to reach the critical current for full compensation of the damping. Similar experiments performed on extended YIG/Pt bilayers did not evidence sizeable modulation of the damping \cite{17hahn,21kelly}, while by monitoring the parametric threshold of SW generation, it was possible to demonstrate a SOT-induced variation of the damping by only up to 7.5\% \cite{lauer:apl2016}.

In the experiment depicted in Fig.\ \ref{fig:yigpt-damping}b, a YIG(20~nm)/Pt(8~nm) bilayer is patterned into a stripe waveguide of width 1$~\mu$m. The bias magnetic field is applied in the plane, transverse to the stripe direction, and an rf line patterned on top of the waveguide allows to excite a SW mode propagating away from the antenna. The SW propagation characteristics are monitored as a function of the dc current injected into the Pt layer by a microfocused Brillouin light scattering setup \cite{evelt:apl2016}. It is observed that the propagation length of the inductively excited SWs can be increased by nearly a factor 10 as the injected current is increased from 0 to 2.5~mA. Therefore, a highly efficient control of magnetic damping can be achieved in this device.

Even if SOT allows a linear modulation of SWs in the sub-critical regime, going above the auto-oscillation threshold, which is marked by a vertical line in the bottom panel of Fig.\ \ref{fig:yigpt-damping}b, leads to a dramatic decrease of the SW amplitude due to nonlinear coupling with other magnon modes, whose amplitudes exponentially grow in this strongly nonequilibrium regime. The same generic phenomenon was observed with stationary SW modes excited by pure spin currents in YIG/Pt microdisks \cite{99collet,100demidov}. As mentioned earlier, these signatures of nonlinear saturation are reminiscent of the well-known physics of FMR in extended films, where short wavelength SW instabilities quickly develop as the excitation power is increased, preventing to achieve large cone angles of uniform precession \cite{Lvov:94}. Interestingly, these nonlinear properties can be adjusted by the perpendicular anisotropy of the film. In this respect, the growth of ultra-thin films of Bi doped YIG exhibiting out-of-plane magnetization and maintaining a high dynamical quality is extremely interesting \cite{soumah:natcom2018}. As a matter of fact, it was demonstrated that their integration in SOT devices could allow the generation of coherent propagating magnons \cite{evelt:pra2018}.

\section{Spin Transport and Manipulation in Magnetic Insulators}

Transport of spin information is possible in many materials, including metals \cite{101johnson}, semiconductors \cite{102lou} and two-dimensional materials such as graphene \cite{103tombros}. In these materials, the conduction electrons mediate the spin flow. However, disturbances in the localized spins can also propagate and carry spin information. Remarkably, spin angular momentum can be transported over distances as large as 40 microns in YIG \cite{2cornelissen} and as far as 80 microns in hematite \cite{3lebrun}. The magnon diffusion length is approximately ten microns at room temperature in YIG \cite{2cornelissen}. When metals are put in contact with magnetic insulators, the spin transport between them can influence the charge transport properties of the conductors.

\subsection{Spin Hall Magnetoresistance}

In magnetic conductors, currents induced by electric fields depend on the orientation of the localized spins. It is the spin-orbit coupling that causes this connection between the electron flow and the orientation of the localized spins.  In isotropic materials, manifestations of this behavior are magnetoresistance and anomalous Hall effect. When the current is along the electric field, the resistance depends on the relative orientation between the current and the localized spins. When the current flows along the electric field, the longitudinal resistivity is 
\begin{equation}
\rho_l = \left( \rho_0 + (\hat{j} \cdot \hat{n})^2 \rho_{amr} \right) \, , 
\label{eq:magnetoresistance}
\end{equation}
where $\hat{j}$ is a unit vector along the current direction and $\hat{n}$ is a unit vector along the direction that describes the preferred spin direction. In ferromagnets, $\hat{n}$ is parallel to the magnetization and in antiferromagnets, $\hat{n}$ is parallel to the staggered order parameter. An out-of-plane orientation of the localized spins can also cause a transverse current. The anomalous Hall effect determines that the transverse current perpendicular to the electric field is governed by a transverse resistivity
\begin{equation}
\rho_t = \rho_{ah} n_z E
\label{eq:anomalousHall}
\end{equation}
where $n_t$ is the transverse component of $\hat{n}$. The magnetoresistance and Hall effects can be used to detect the orientation of the localized spins \cite{67nakayama,68aqeel,Chen:PRB2013,Hou:PRL2017}. 

In magnetic insulators, there are no charge currents. Naturally, there are then neither an associated magnetoresistance nor an anomalous Hall effect. However, as we have discussed, spins can propagate across metal-magnetic insulators interfaces when they are perpendicular to the spins in the magnetic insulators. Consequently, in layered metal-magnetic insulator systems, spin transport within the metallic regions will be affected by the transverse spins that flow into the magnetic insulators. The spin Hall effect generates spin currents that will experience this additional interfacial transverse decay of spins. The reciprocal effect, the inverse spin Hall effect, generates a charge current from the spin current. There is, therefore, a second-order effect in the spin Hall angle on the charge transport properties. First, the spin Hall effect generates a transverse spin current from the primary charge current that, in turn, causes a change in the charge current via the inverse spin Hall effect. These mechanisms imply that the charge resistance will be affected by the spin transport properties and, hence, via the spin decay at the metal-magnetic insulator interface. 

The result of combination of the spin Hall effect, spin-transfer, and inverse spin Hall effects is that the transport properties in layered metal-magnetic insulators are no longer isotropic. The current depends on how the metal is attached to the magnetic insulator. Such a magnetoresistance in heterostructures of metals and insulators is dubbed a spin Hall magnetoresistance since it heavily relies on the spin Hall effect \cite{17hahn,67nakayama,Chen:PRB2013}. 

In a layered metal-magnetic insulator structure, where the $x$ and $y$ axes are in-plane, and the $z$ axis is perpendicular to the metal-insulator interface, the anisotropy of the current is richer than the behaviour of Eqs.\  \ref{eq:magnetoresistance} and \ref{eq:anomalousHall}. When the current flows along the $x$ direction, the longitudinal and transverse resistivities become \cite{67nakayama,68aqeel,Chen:PRB2013}
\begin{equation}
\rho_l = \rho_0 + \rho_1 (1-m_y^2) \, ,   
\label{eq:smr} 
\end{equation}
and 
\begin{equation}
\rho_t = \rho_1 m_x m_y + \rho_2 m_z \, . 
\label{eq:smrHall}
\end{equation}
The spin Hall magnetoresistance represented by the coefficient $\rho_1$ in Eq.\ \ref{eq:smr} is quadratic in the spin Hall angle since the underlying mechanism is the combination of the spin Hall effect and the inverse spin Hall effect \cite{Chen:PRB2013}. Furthermore, it depends on dissipative part the transverse conductance (the real part of the "mixing conductance" \cite{Brataas:PRL2000}) that describes the Slonczewski spin-transfer between the metal and the magnetic insulator that influences the spin accumulation in the metal. There is also a more conventional anomalous Hall contribution to the transverse resistance of Eq.\ \ref{eq:smrHall}, represented by the coefficient $\rho_2$. This term depends on the reactive part of the transverse conductance (the imaginary part of the "mixing conductance" \cite{Brataas:PRL2000}) \cite{Chen:PRB2013} that governs the the magnetic proximity effect in the normal metal.

It is also possible to understand the anisotropy of the longitudinal and transverse resistivities from symmetry arguments. In bilayers of metals and magnetic insulators, the reduced symmetry along the axis normal to the interface leads to a qualitative change of the properties of the resistivities. Experimentally, not only the qualitative behavior of the spin Hall magnetoresistance, but also its magnitude agrees well with the expression \cite{67nakayama,Chen:PRB2013}.  This agreement shows that the combination of the spin Hall effect, the spin-transfer, and the inverse spin Hall effect describe the essential physics well \cite{Chen:PRB2013}. The spin Hall magnetoresistance is an interfacial effect on a bulk transport property. It is, therefore, small and only relevant in thin metallic films. Nevertheless, it is essential because it is a new way to detect the spin order in both ferromagnets and antiferromagnets. 

\subsection{Nonlocal Devices}

In nonlocal devices, the electrical current flowing in the metallic wire can affect the magnon population inside the magnetic insulator in two ways. Spin-transfer is the first mechanism and can be coherent or incoherent. Coherently, spin-polarized currents generated inside the metallic wire due to spin-orbit effects transfer to the adjacent magnetic insulator layer and produce a torque on the magnetization. Incoherently, at elevated temperatures, there can be a spin flow parallel to the magnetization by thermal magnons. Joule heating is the second mechanism. The ohmic dissipation in the metallic wire locally increases the magnetic layer temperature, which is in thermal contact, and, correspondingly, there is a local increase in the number of thermal magnons.

\begin{figure}[htbp]
\includegraphics[width=0.99\columnwidth]{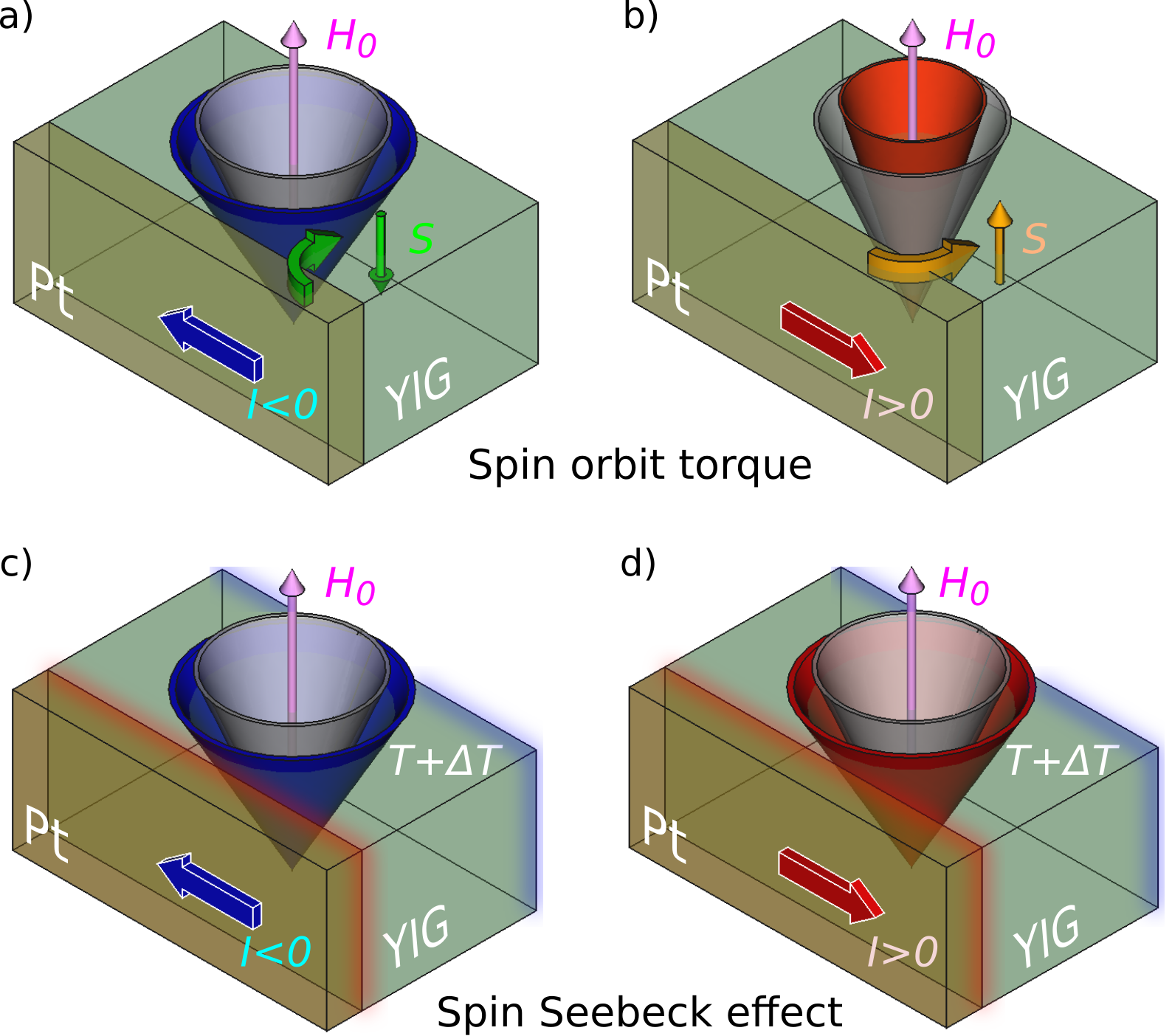}
%\vspace{-8pt}
\caption{Schematic illustration of spin-transfer processes between a magnetic insulator and an adjacent metallic layer where an electric current circulates. The first row shows the principle of SOT, where the electron trajectory inside the normal metal is deflected towards the interface depending on the spin polarization of the electron. The second row shows the principle of SSE, where the temperature rise associated with Joule heating changes the distribution of thermal magnons.  Changes in the amplitude of the magnon mode is illustrated as a deviation of the cone angle from the nominal occupation (thermal population at the device temperature).}
%\vspace{-8pt}
\label{fig:sotseebeck}
\end{figure}

To separate these two contributions, we rely on the symmetry of the signal with respect to the current or field polarity. We first concentrate on the current symmetries. In the case of SOTs, the electron trajectory inside the normal metal is deflected towards the interface depending on the spin polarization of the electron. For metals such as Pt, where the spin Hall angle  \(  \theta _{sh} \)  is positive, the deflection follows the right-hand side rule as shown schematically in Fig.\ \ref{fig:sotseebeck} a) and b) for both positive and negative current. The result is an inversion of the polarity of the outward spin current when the electrical current direction is reversed. The net effect is an opposite change in the magnon population as shown in Fig.\ \ref{fig:sotseebeck} a) and b). The nominal occupation (thermal population at device temperature, \( T\)) of a particular spin-wave mode with eigen-value \( \omega_m \) is non-zero and it writes  \( k_B T / (\hbar \omega_m) \). The corresponding amplitude is depicted as a gray cone in Fig.\ \ref{fig:sotseebeck}. If the spin is injected parallel to the magnetization direction, then the cone angle will decrease (magnon annihilation, Fig.\ \ref{fig:sotseebeck} b), while if the spin direction is opposite to the magnetization direction, then  the cone angle will increase (magnon creation, Fig.\ \ref{fig:sotseebeck} a). Thus, an important signature of SOT is that the signal is odd with respect to the current polarity. In contrast, a change of the magnon population produced by thermal effects is even in current, since the origin is Joule heating which is proportional to $I^2$. This distinction between even and odd symmetries with respect to the current polarity translates into a signal appearing in different harmonics when performing lock-in measurements \cite{2cornelissen}. In the case of SOT, which is odd in current, the signal is captured by the 1st harmonics, while in the case of the spin Seebeck effect (SSE), which is even in current, the signal is captured by the 2nd harmonics. Ref.\ \cite{Goennenwein:APL2015} also found the symmetry expected for a magnon spin-accumulation-driven proces and confirmed the results in Ref.\ \cite{2cornelissen}.

%%%%%%%%%%%%%%%%%%%%%%%%%%%
%%%%%%%%%%%%%%%%%%%%%%%%%%%%

These symmetries concerning the current direction and the symmetries related to the magnetization direction have a direct correspondence. The magnetization is, in the case of YIG, determined by the external magnetic field orientation. However, we also need to take into account that the signal polarity is inverted when the magnetization underneath the detector reverses. Thus, the net effect is that the SOT signal is even, while the SSE signal becomes odd with respect to the field polarity. If one observes the azimuthal angular dependence, the even signal follows a cos behavior, while the odd signal follows a $\cos^{2}$ behavior. Since the torque is exerted on the transverse magnetization (the oscillating part of the magnetization), the effect is maximum when the saturation magnetization is parallel to the injected spin direction, or, in other words, perpendicular to the current flow in the Pt, as first demonstrated in Ref. \cite{2cornelissen}.

\begin{figure}[htbp]
\includegraphics[width=0.99\columnwidth]{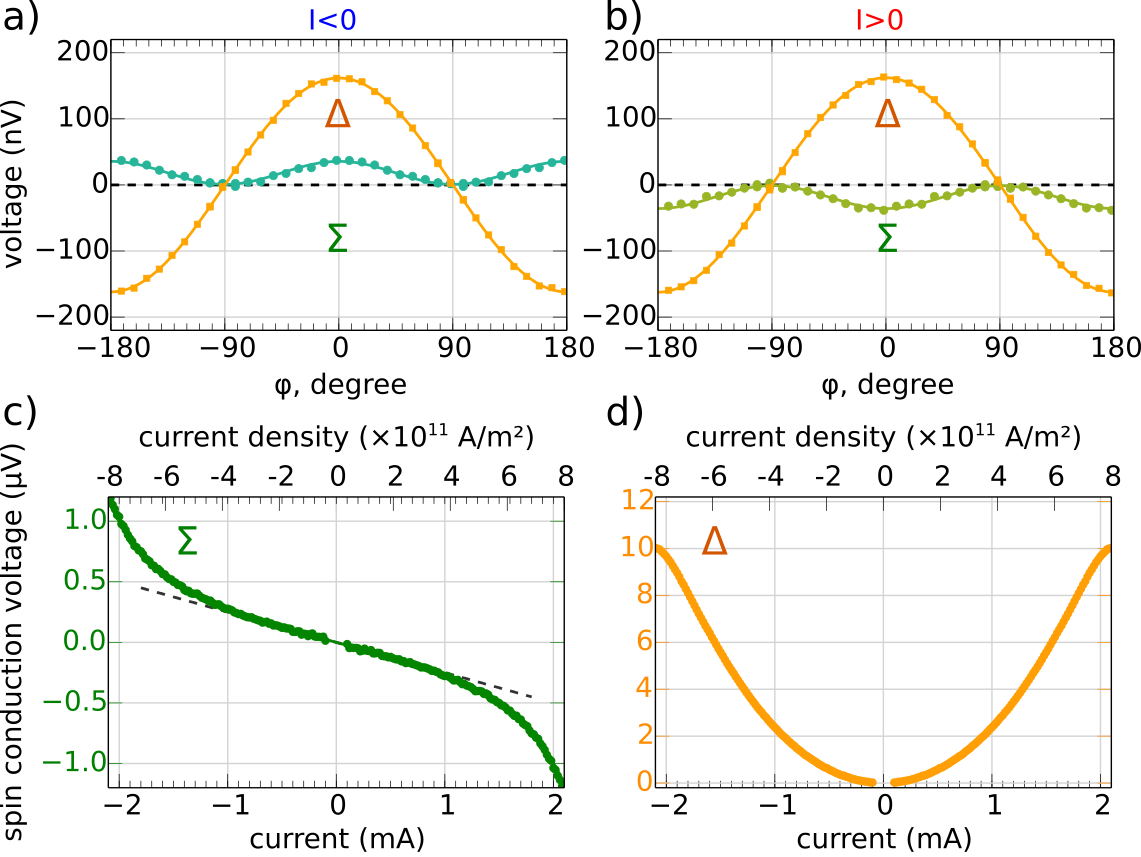}
%\vspace{-8pt}
\caption{a,b) Azimuthal dependence of the anisotropic voltage produced in a nonlocal device, which can be separated into two components : $\Sigma$, even in magnetic field polarization, and $\Delta$, odd in magnetic field polarization. c,d) Current dependence of respectively $\Sigma$  and $\Delta$ over a large amplitude of current density span measured on a 19nm thick YIG film. Panels a,b) are adapted from Refs.\ \cite{105thiery} and c,d) from Ref.\ \cite{104thiery}.}
%\vspace{-8pt}
\label{fig:measurements}
\end{figure}

Importantly, in the nonlocal transport experiment, the voltage drop produced on the detector side can also have an electrical origin \cite{104thiery}. Among the electrical effects are ohmic loss, the ordinary Hall effect, the thermoelectric effect, and the thermal Hall effect. These electrical contributions can originate from thermally activated conduction inside the YIG layer caused by dopants or grain boundaries \cite{104thiery} or spurious conduction channels inside the substrate or the presence of a capping layer. These electrical and thermal effects must be disentangled from pure spin effects. To separate the pure spin contribution, Ref. \cite{2cornelissen} proposed solely considering the anisotropic part of the transport as one varies the orientation of the in-plane magnetization. Among these anisotropic contributions, as emphasized above, one should thoroughly distinguish the ones that are even with respect to the magnetic field or current polarity from those that are odd. In Ref. \cite{2cornelissen} the authors, using the 1st and 2nd harmonic output of a lock-in amplifier, relied on the current symmetry to extract the SOT and SSE. These contributions can also be extracted by directly measuring the voltages obtained under the two possible polarities of the externally applied magnetic field, and constructing respectively a signal sum ($\Sigma$) and a signal difference ($\Delta$) from these measurements \cite{105thiery}. This process is summarized in Fig.\ \ref{fig:measurements} a) and b) and the corresponding current evolution is shown in Fig.\ \ref{fig:measurements} c) and d) for both $\Sigma$  and $\Delta$  when measured on a 19 nm YIG film. Importantly, the\ part that is even with respect to the magnetic field polarity ($\Sigma$) is odd with respect to the current polarity, while the part that is odd with respect to the magnetic field polarity ($\Delta$) is even with respect to the current polarity.  These results are exactly the expected symmetries of SOT effect and SSE, as explained earlier.

%%%%%%%%%%%%%%%%%%%%%%%%%%%
%%%%%%%%%%%%%%%%%%%%%%%%%%%%

The features in these nonlocal voltages also appear in the local voltage, which can be viewed as self-detection of the ISHE voltage on the injector side. In the case of the $\Sigma$-signal, the produced effect is the so-called spin Hall magneto-resistance \cite{17hahn,67nakayama}. As explained previously, the signal is maximum when the magnetization is perpendicular to the current flow inside the Pt and minimum when the magnetization is parallel to the current flow. Experimentally, however, this formulation seems at a first glance counterintuitive since the total resistance appears to drop when the magnetization is perpendicular to the current flow. This result occurs because the drop in the voltage produced by SOT is negative for positive current (and vice versa) and thus appears as a 'negative' resistance effect (see Fig.\ \ref{fig:measurements}c). This drop is a direct consequence of the Hall origin of the $\Sigma$-signal. Contrary to ohmic loss, where the voltage drops along the current direction, for the ISHE, the Hall voltage increases along the current direction \cite{104thiery}. 

In the same way, a local-SSE signal \cite{106schreir} should appear in the device synchronously with the nonlocal $\Delta$-signal mentioned previously. This local $\Delta$-signal is synchronous with\ the  $\Delta$ observed for a very small gap, but negative with respect to the sign, which just means that the vertical thermal gradient is negative, i.e. that the YIG$\vert$Pt interface is hotter than the YIG$\vert$GGG interface. The current dependence of this local-SSE should be quadratic (as $I^{2}$, Joule heating) in current. Note that this behavior has exactly the same signature as the uniaxial magneto-resistance \cite{107avci}. In this respect a resistance that is linear in current and change signs when the current is reversed is equivalent to a voltage that varies quadratically with $I^{2}$, which is usually interpreted as a Joule heating signature.

\begin{figure}[htbp]
\includegraphics[width=0.99\columnwidth]{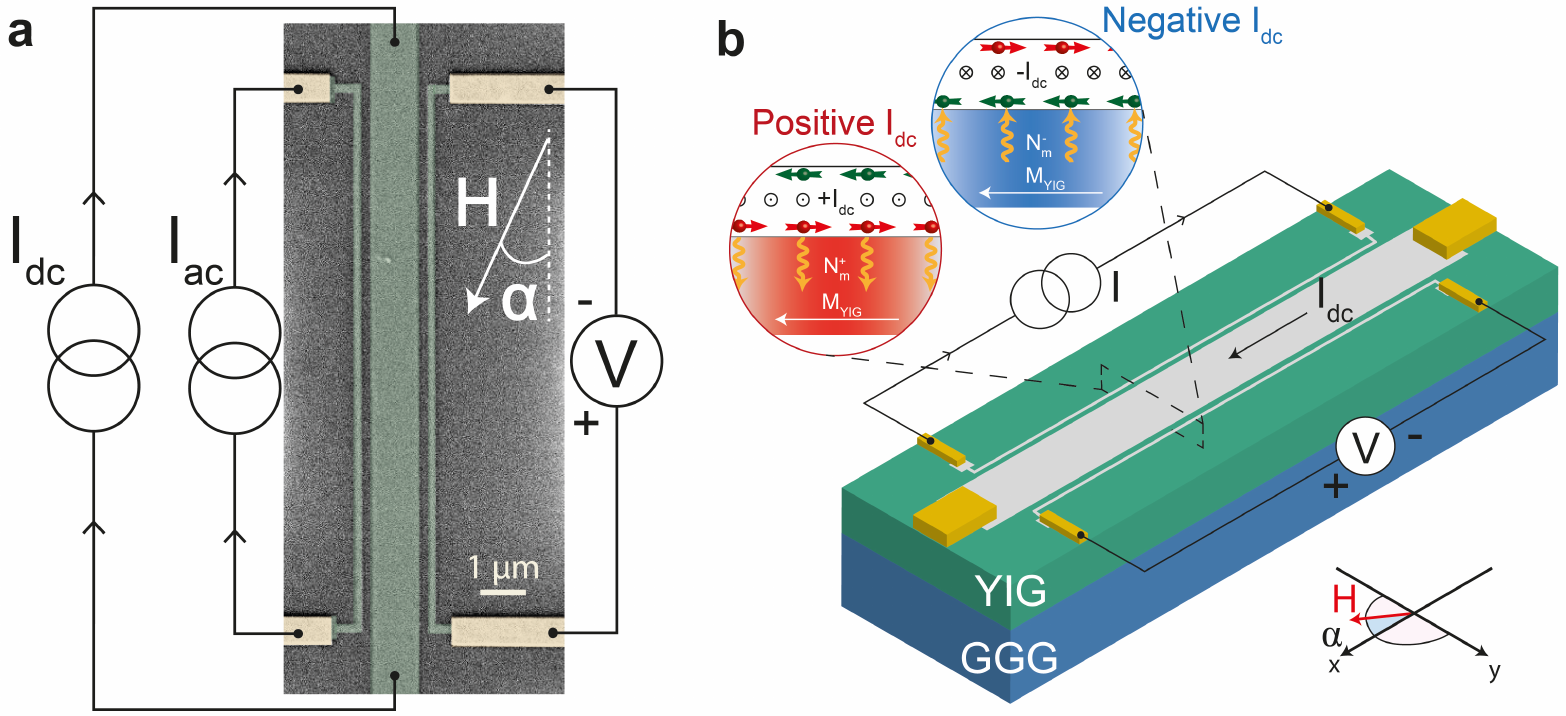}
%\vspace{-8pt}
\caption{Principle of electrically injected and detected nonlocal magnon transport and a magnon spin transistor. An AC current thought the Pt injector strip generates an AC magnon accumulation via the spin Hall effect. These magnons diffuse to the Pt detector strip, generating a charge voltage via the ISHE. An intermediate gate electrode enables the control of the magnon density, and the magnon conductivity. In this way the nonlocal signal can be modulated by a dc current through the gate electrode. The figure is from Ref.\ \cite{116cornelissen} }
%\vspace{-8pt}
\label{fig:nonlocal}
\end{figure}

The incoherent magnon transport in insulating materials such as YIG is dominated by thermal magnons whose number overwhelmingly exceeds the number of other modes at finite temperature.  This phenomenon is demonstrated in the SSE \cite{108uchida}, where  a transverse voltage in a Pt electrode fabricated on a YIG layer develops as a result of thermally induced magnon spin transport. This process can also be reversed \cite{2cornelissen,109zhang} and can be used for electrically driven magnon spin injection. In the nonlocal geometry of Fig.\ \ref{fig:nonlocal}, a charge current through the Pt injector strip (left)\ generates an electron spin accumulation at the Pt/YIG interface via the spin Hall effect.  Exchange processes across the interface result in a magnon spin accumulation and a non-zero magnon chemical potential. This potential drives magnon diffusion and at the detector electrode (right), the resulting nonzero magnon chemical potential results in an electronic spin accumulation, which drives an electron spin current and is partially converted into an electrical voltage via the ISHE.\  By varying the spacing between the injector and detector electrodes, a typical magnon spin relaxation length of 10 microns at room temperature could be determined. This result was confirmed by simultaneously measuring the effects of electrical as well as thermal magnon injection.\  The SSE under the injector electrode gave rise to the latter effect. The SSE has also been used to control damping \cite{110jungfleisch} and even to generate auto-oscillations of magnetization \cite{111safranski}.

%%%%%%%%%%%%%%%%%%%%%%%%%%%
%%%%%%%%%%%%%%%%%%%%%%%%%%%%

This and other experiments \cite{Ganzhorn:APL2016,113wu,114li} confirmed that in addition to driving a magnon spin current by a temperature gradient, the magnon chemical potential plays a crucial role in driving magnon currents  in magnetic insulators. 

The universality of this nonlocal technique has been recently shown in the study of thermal magnon transport in the most ubiquitous antiferromagnet $\alpha$ Fe\textsubscript{2}O\textsubscript{3}, where typical magnon spin relaxation lengths of 10 microns\ \ were  observed at temperatures of 200 K \cite{3lebrun}. An external magnetic field can control the flow of spin current across a platinum-hematite-platinum system by changing the antiferromagnetic resonance frequency. The spin flow is parallel to the Neel order. Magnon modes with a frequency of tens of GHz, or as large as 0.5 THz can carry spin information over microns. Importantly, this result demonstrates the suitability of antiferromagnets in replacing currently used components. Antiferromagnets can operate very quickly and are robust against external perturbations. Speed is not detrimental to the spin diffusion in these materials, which opens the possibility of developing faster devices. As we have discussed, AFIs are more prominent than FIs. These developments open the door towards exploring a wider class of materials with a richer physics.

\begin{figure}[htbp]
\includegraphics[width=0.99\columnwidth]{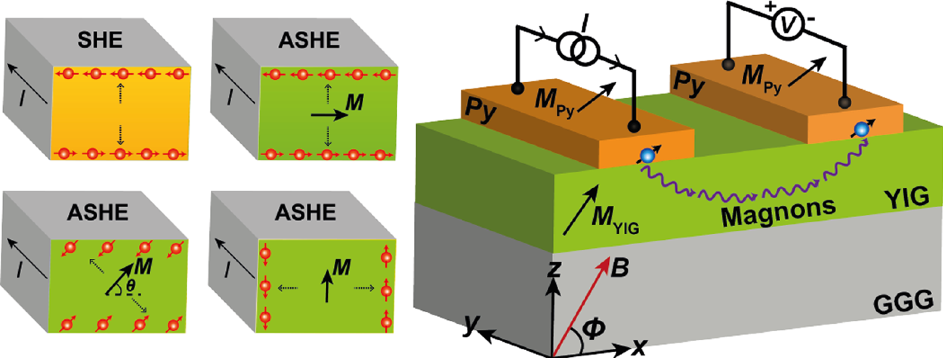}
%\vspace{-8pt}
\caption{(Right panel) Efficient and polarization controlled injection and detection of magnon spins via the anomalous spin Hall effect and its inverse. (Left panel) (a) Magnon spin injection and detection via the spin Hall/inverse spin Hall effect. At the metal/YIG interface the spins have an in-plane orientation. (b,c,d)  The anomalous spin Hall effect in a ferromagnetic produces spins which are oriented parallel to the magnetization. This allows magnetization control of the injected magnon spins. (from \cite{115das})}
%\vspace{-8pt}
\label{fig:nonlocalsheferro}
\end{figure}

In addition to magnon spin injection and detection using the spin Hall and inverse spin Hall effect, it was discovered that ferromagnetic metals, such as Py, can also inject  and detect magnon spins effectively into a magnetic insulator via the anomalous spin Hall effect and its inverse \cite{115das}.  Unlike the SHE, which can only produce in-plane polarized spins, a charge current though a ferromagnet can also produce an anomalous spin Hall effect (ASHE), which generates a spin current perpendicular to the charge current and magnetization direction,  but where the spins are oriented parallel to the magnetization,  see Fig.\ \ref{fig:nonlocalsheferro}. Since the latter can be controlled by a magnetic field or by other means, this allows the injection and/or detection of magnon spins which have an out-of-plane polarization component.  The coexistence of this mechanisms for (magnon) spin injection/detection with ferromagnetic electrodes is discussed in \cite{Amin:PRB2019}.

Ref.\ \cite{116cornelissen} reported the realization of a spin transistor by inserting a third "gate"  electrode in between the magnon injector and detector \cite{116cornelissen}; see  Fig.\ \ref{fig:nonlocal}. A dc current flows through the intermediate electrode. Depending on the current direction, the magnon density in the YIG below is either enhanced or reduced, leading to a modulation of the magnon conductivity and the nonlocal signal by several percent. The experiments were modelled with a finite element model, which included the modification of the YIG magnon spin conductance by the magnon injection of the control electrode. The magnon spin conductance was typically found in the order of the spin conductance of a poor metal like Pt. The modelling showed that at the maximum control current, the magnon chemical potential was around $10 \mu eV$, which resulted in a modification of the magnon density of a few percent.

In a similar geometry, but by making the YIG thinner and using higher currents, Ref.\ \cite{Wimmer:PRL2019} realized full control of the magnetic damping via spin-orbit torques. When the current flowing through the intermediate electrode exceeds a critical value, the significantly reduced damping gives rise to a substantial increase of the magnon conductivity by almost two orders of magnitude. Ref.\ \cite{Wimmer:PRL2019} discusses three possible scenarios for the massive increase in the magnon conductivity, 1) injection of magnons in broad frequency spectrum compensates the damping without developing a coherent state, ii) compensation of magnetic damping leads to coherent magnetic auto-oscillations, and iii) the magnon system form a BEC with a macroscopic population of the ground state. In a control experiment, when microwaves coherently drive the magnetization, the spin conductivity decreases, exhibiting the opposite behavior. This measurement shows that it is the compensation of the magnetic damping that is responsible for the substantial enhancement in the spin conductivity rather than the coherent magnetization dynamics. The experiment opens news ways to actively control the magnon transport from regimes where the spin resistance is significantly reduced or vanishes to larger and finite values.

Although the role of thermal magnons dominates the observed behavior at a low current, the nonlocal $\Sigma$-conductivity can exhibit nonlinear behavior at a large current density due to distortion of the magnon distribution produced by SOT. SOT is a process whose efficiency increases with decreasing magnon frequency. This process thus favors low energy magnons present near the spectral bottom of the magnon manifold, the region of so-called magnetostatic waves. In the high out-of-equilibrium regime, these low-energy magnons can efficiently thermalize between themselves through the potent magnon-magnon interaction, whose strength increases with the magnon population. Describing the quasi-equlibrium state by a nonzero chemical potential \cite{4demokritov} and an effective temperature \cite{118serga} is insightful. This state forms the so-called subthermal magnons, a term coined in order to distinguish them from the thermal magnons, from which they are effectively decoupled, as under the intensive parametric pumping one can reach a state where the effective temperature of subthermal magnons exceeds the real temperature characterizing the thermal magnons by a factor of 100. 

For nonlocal spin transport device, this modification in the magnon distribution implies a change in the spin transport properties. For very large SOT, the conductance becomes dominated by the propagation properties of the subthermal magnons, which leads to an increase of the $\Sigma$-signal at high current since these magnons have a higher characteristic propagation distance than thermal magnons. This distortion is visible as a gradual deviation of the spin conductance from a purely linear transport behavior at large I, as observed in Fig.\ \ref{fig:measurements} c). A crossover current (here approximately 1.4 mA) from linear to nonlinear regime \cite{105thiery} is observed. The deviations are subcritical spin fluctuations, and the crossover current occurs below the critical current for damping compensation of coherent modes. The slow rise, which is in contrast with the sudden surge of coherent magnons observed at the critical threshold in confined geometries \cite{119hamadeh}, should thus be considered as a signature of the phenomena of damping compensation in open geometries.

Two interesting consequences arise from this phenomenon. The first one is that these effects imprints some nonlinear properties on these devices, which could be advantageously exploited to produce harmonic generation, mixing or cut-off effects in insulatronics. An additional consequence could also emerge from the specific nature of the subthermal magnons. For example, while the transport of thermal magnons is difficult to control due to their relatively high energies, the subthermal magnons could be efficiently controlled by relatively weak magnetic fields.

\section{Future Perspectives}

Spin insulatronics is an emerging field. In magnetic insulators, the ultralow dissipation facilitates brand new exciting possibilities for the exploration of novel and rich physical phenomena. The development of new materials or material combinations will also empower future developments. Let us explore possible improvements in materials and how coherent spin dynamics can open novel avenues for spin transport.

\subsection{Materials and Interfaces}

The material properties of magnetic insulators and their interfaces are essential. We will first discuss the interface properties. The efficiency of the injection of spins into magnetic insulators from metals depends on the interfaces. The detection efficiency depends on the same parameters. Devices require injection and detection of spins. The performance is, therefore, quadratic in the interface spin transparency, which is an essential factor. Measurements on YIG/platinum systems and YIG/gold systems \cite{70sandweg,71sandweg,72vilela,73rezende,74azenvedo,75burrowes,Heinrich:PRL2011} have established that the efficiency substantially varies with the preparation technique even for the same material combinations \cite{121jungfleisch}. While many experiments show that there can be a robust coupling between FIs and metals, there has been less exploration of AFIs. Pt couples to hematite as strongly as to insulating ferrimagnets, such as YIG \cite{3lebrun}. For antiferromagnets, determining how the spins at the two sublattices couple, possibly in different and unusual ways, to the spins in the metals depending on the crystal structures and interface directions would also be of interest. Experimental demonstration of spin-pumping from antiferromagnets in direct contact with metals is desirable \cite{122johansen}. More precise insights into the electron-magnon coupling could enable a stronger and different control of the spin excitations in AFIs. Better spin injections, possibly the use of topological insulators in combination with magnetic insulators, could also increase the efficiency of devices.

In the bulk of magnetic insulators, reduced damping and anisotropy control are essential. Oxide materials offer a broad choice with numerous possible substitutions. Past efforts have mostly concentrated on spinels. Spinels are minerals that crystallize in the cubic form. Some candidates among the spinel ferrites are NiZnAl-ferrite\textsuperscript{1} ($ \alpha $ = 3 10\textsuperscript{-3} in thin film form) and the magnesium aluminum ferrite MgAl\textsubscript{0.5}Fe\textsubscript{1.5}O\textsubscript{4} ($ \alpha $ = 1.5 10\textsuperscript{-3\  }in 10 nm thick films)\textsuperscript{2}. Recent developments in pulsed laser deposition techniques give access to a new class of epitaxial thin films with improved dynamic properties. Illustrative examples are manganite materials such as LSMO \cite{123flovik}, with a reported low damping value. Among the oxide materials studied, apart from garnets, the other compounds that stand out are hexaferrites, of specific interest in the 1980s. Of particular interest are the strontium hexaferrites \cite{124song}, Ba-M hexaferrite \cite{124song}, and zinc lithium ferrite \cite{125song}, where the FMR linewidth can be as low as 30 MHZ at 60 GHz in thin films, making\ them strong contenders for excellent spin conductors.   Antiferromagnets comprise the majority of magnetically ordered materials. AFIs commonly occur among transition metal compounds, where the interaction between the magnetic atoms is indirect (super exchange), e.g., through oxygen ions as in hematite (Fe\textsubscript{2}O\textsubscript{3}), nickel oxide (NiO), cobalt oxide (CoO) or chromium oxides (Cr\textsubscript{2}O\textsubscript{3}).\  Fluorides, such as MnF\textsubscript{2}, are also good potential candidates. In hematite, there long-range spin transport across 80 microns has been demonstrated \cite{3lebrun}. With a higher level of purity, we expect to see spins transport across microns in the broader class of AFIs. We anticipate further demonstrations and investigations of long-range spin transport and manipulation in these wide ranges of materials.

\subsection{Condensation and Superfluidity}

Spin-torque oscillators generate sustainable output ac outputs from dc inputs \cite{126silva,127kim}. In ferromagnets, these oscillators utilize the spin-transfer or spin-orbit torque to evolve into a steady-state oscillation of the magnetization that in turn generates the output signal via magnetoresistance effects. The principle to realize persistent oscillations in ferromagnets is as follows. The spin-transfer torque enhances or reduces the dissipation depending on the current direction. However, in dedicated geometries, the dependence on the precession cone angle can differ from that of the Gilbert damping. Therefore, for one current polarity, the spin-transfer torque compensates the Gilbert damping at an angle where steady-state oscillations occur. In FIs, the reduced dissipation rate reflected in the smaller Gilbert damping constant can facilitate spin-torque oscillators at lower applied currents. We can speculate that insulators might provide new ways to synchronize oscillators, thus producing a desired larger output signal. Spin-torque oscillators at much higher, i.e. THz, frequencies obtained by using antiferromagnets can be envisioned \cite{128cheng,129khymyn,130sulymenko,Shen:APL2019}. In antiferromagnets, it is the steady-state oscillation of the Néel order that generates the output signal. Typically, the antiferromagnetic resonance frequency is much higher than the resonance frequencies in ferromagnets. This feature enables the creation of THz electronics that can boost the field of spintronics and other branches of high-speed electronics. In AFIs, the small damping rate implies a reduced current to reach the auto-oscillation threshold, which might make the first experimental demonstration easier to carry out.

\begin{figure}[htbp]
\includegraphics[width=0.6\columnwidth]{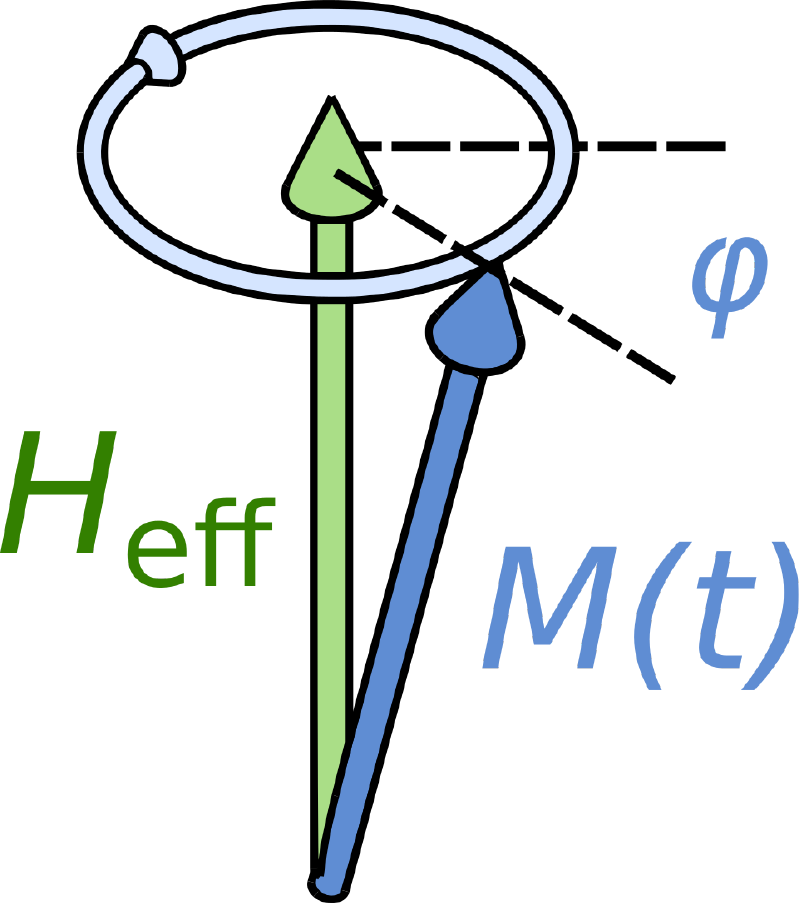}
%\vspace{-8pt}
\caption{Semiclassical picture of magnon condensation. Coherent dynamics with a phase $\phi$ exists. The transverse components of the magnetization precess with the ferromagnetic resonance frequency $\omega$.}
%\vspace{-8pt}
\label{fig:condensation}
\end{figure}

Super spin insulatronics is the ultimate quantum limit of magnon condensation and spin superfluidity. Traditionally, in magnetic insulators, the utilization of spin phenomena are utilized via semiclassical SWs \cite{7kruglyak,Serga:JPD2010}. Beyond this regime, at a sufficient density, magnons condense into a single Bose quantum state. In ferromagnets, magnon BEC manifests itself by a phase-coherent precession of the magnetization. Let us consider a ferromagnet in which the magnetization is homogenous in the ground state and the lowest energy excitations are also homogenous. Semiclassically, a unit vector along the magnetization represents the magnetic state. At equilibrium, this vector along an axis that we take to be the longitudinal  \( z \)  direction. Condensation is represented by a small deviation of magnetization in the transverse directions,
\begin{equation}
m_{+}=m_{x}+m_{y}=a \exp (i \omega t+\varphi) \, , 
\end{equation}
where the real number  \( a \)  is an amplitude,  \(\omega\)  is the ferromagnetic resonance precession frequency, and  \(\varphi\)  is the phase of the condensate. We sketch the magnon condensation in Fig.\ \ref{fig:condensation}. The reduction of the unit vector of the magnetization along the longitudinal direction  \(  \delta m_{z}=a^{2}/2 \)  is proportional to the number of magnons. The condensation is manifested in the larger magnon population at the energy minimum of the magnon bands as compared to that described by the Bose-Einstein distribution or, at high temperatures, than the Rayleigh-Jeans distribution \cite{131ruckriegel}. In antiferromagnets, magnon condensation is similar to the phenomenon in ferromagnets, but it is the Néel field that undergoes phase-coherent precession rather than the magnetization.

%%%%%%%%%%%%%%%%%%%%%%%%%%%
%%%%%%%%%%%%%%%%%%%%%%%%%%%%

In thin ferromagnetic films, the dipole-dipole interaction dramatically changes the magnon dispersion. When the ground state magnetization is in-plane, the dispersion is anisotropic, and the energy minimum is at a finite wave-vector. The magnons will then condense around this energy minimum. Importantly, Ref.\ \cite{4demokritov} observed spectroscopically generated magnon condensates in a thin film of a ferrimagnetic insulator at room temperatures. The first observations were that the occupation of the lowest energy state was considerably higher than the surrounding states than expected according to the Bose-Einstein distribution.  Subsequent studies showed that the condensate is coherent \cite{Demidov:PRL2008}. In these experiments, parametric pumping by microwave fields parallel to the equilibrium magnetization created a large number of out-of-equilibrium magnons. Four-magnon interactions cause some of these magnons to relax their energies towards the minimum energy. When the pump is turned off, evaporative cooling channels some magnons to much higher energies as compared to the vicinity of the energy minimum and the condensate forms. The typical time-scale related to the magnon-magnon relaxation is shorter than the spin relaxation time associated with the spin-lattic interaction. The stability of the condensate requires that the interaction between the magnons is repulsive \cite{Dalfovo:RMP99,Pitaevskii:2003}. This criterion can seems at odds with the experimental results   \cite{Tupitsyn:PRL2008} because established theories predict that the interaction between the condensed magnons is attractive \cite{Lvov:94,Gurevich:CRC96}. It was recently experimentally shown that the Bose-Einstein condensate is stable since the effective interaction between the magnons in spatially inhomogeneous systems is repulsive \cite{Borisenko:NatCom2020}. A possible source of the repulsive interaction is the dipole field arising from the inhomogenous condensate density \cite{Borisenko:NatCom2020}. Recently, a new way to realize Bose-Einstein condensation by rapid cooling was also demonstrated \cite{Schneider:NatNano2020}.

NMR-induced condensation and superfluidity in antiferromagnets have also been reported \cite{132bunkov,133bunkov}. Crucial for spin insulatronics, a hypothesis that we can electrically control magnon condensation via spin-transfer and spin-pumping both in ferromagnets \cite{95bender,134bender} and antiferromagnets \cite{135fjarbu} has been proposed. These measurements and theoretical suggestions imply that coherent quantum phenomena that utilize magnons could possibly be demonstrated in the future. Eventually, it might become possible to use these aspects in devices without the need for complicated cooling devices. 

\begin{figure}[htbp]
\includegraphics[width=0.99\columnwidth]{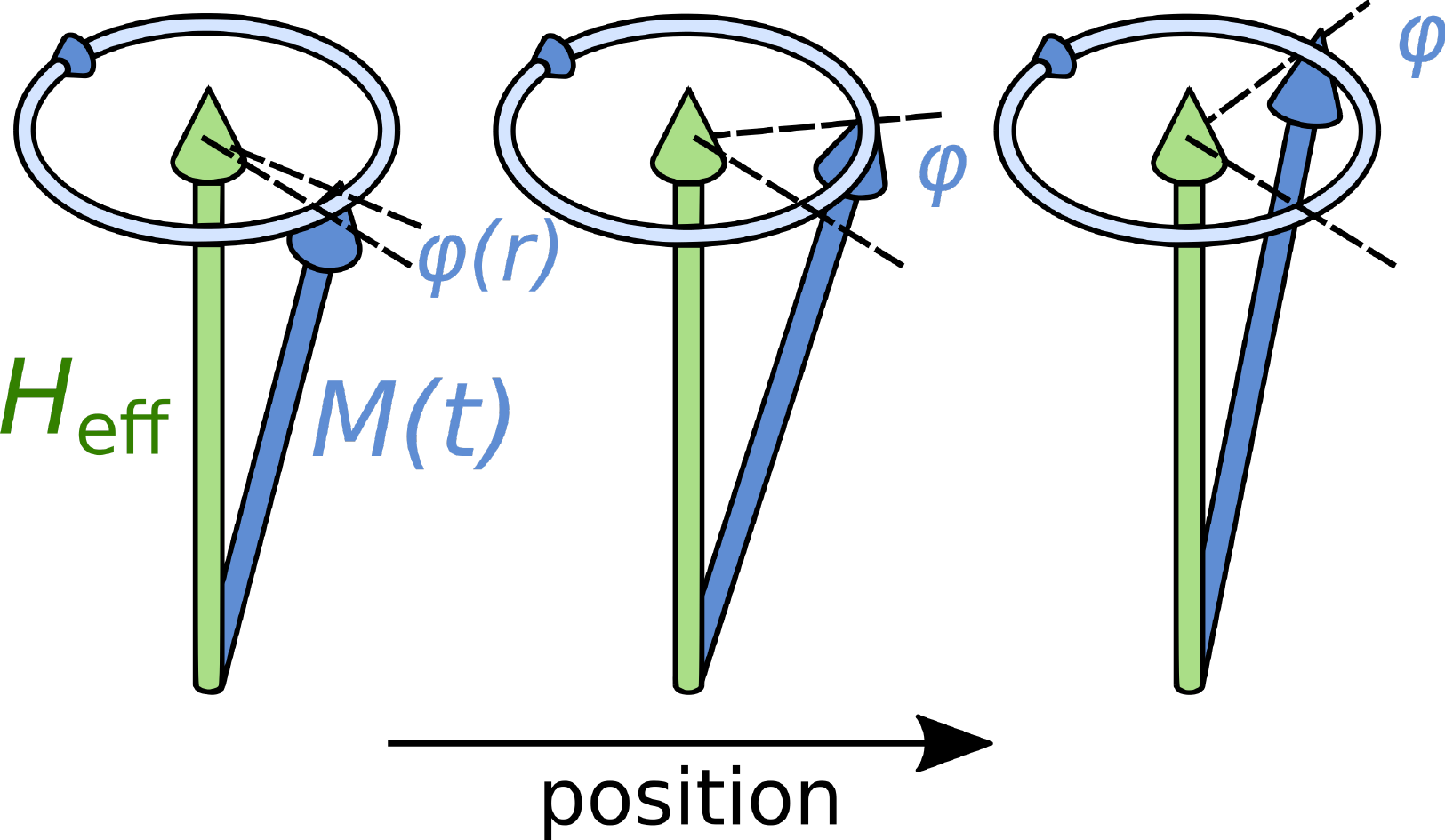}
%\vspace{-8pt}
\caption{Spin superfluidity. The phase of the condensate exhibits spatial variation. The dissipationless spin supercurrent is proportional to the gradient of the phase.}
%\vspace{-8pt}
\label{fig:superfluidity}
\end{figure}

Superfluidity is a dissipationless flow governed by the gradient of the condensate phase. In the absence of spin-relaxation, an analogy exists between the spin dynamics in planar magnetic systems and the hydrodynamic behavior of ideal liquids \cite{5halperin}. These concepts have been subsequent extended, dubbed spin superfluidity \cite{136sonin,137sonin}. When magnons condense, the phase of the order parameter equals the phase of the corresponding semiclassical precession angle. The spin current is proportional to the spatial variations of the precession angle \(\varphi\).
\begin{equation}
j_{s}=A \nabla \varphi \, , 
\end{equation}
where \(A\) is a constant related to the spin-stiffness. Fig.\ \ref{fig:superfluidity} presents the physics of spin superfluidity.

%%%%%%%%%%%%%%%%%%%%%%%%%%%
%%%%%%%%%%%%%%%%%%%%%%%%%%%%

Electrical injection of superfluid transport is possible via spin-transfer and spin-pumping \cite{138takei}. In other words, the bulk superfluid spin current is supplemented with boundary conditions for the injection and detection of the spin currents. These boundary conditions depend on the contacts via the transverse (or "mixing" ) conductance. The physics is similar to the case of the conductance of normal metal-superconductor-normal metal systems that also depends on the contact interfaces between the normal metals and the superconductors. However, differences exist between the case of spin superfluidity and dissipationless charge transport. Although no free carriers exist in insulators, spins, nevertheless, interact with lattice vibrations, which causes dissipation. As a result, in a simple one-dimensional geometry, in a metal-magnetic insulator-metal system, the ratio between the emitted spin current and the spin accumulation bias becomes
\begin{equation}
\frac{j_{s}}{\mu_{s}}=\frac{1}{4 \pi} \frac{g_{L} g_{R}}{g_{L}+g_{R}+g_{\alpha}} \, , 
\end{equation}
where  \(g_{L}\)  is the transverse ("mixing" ) conductance of the left metal-magnetic insulator contact,  \( g_{R} \)  is the transverse ("mixing" ) conductance of the right magnetic insulator-metal contact, and  \( g_{ \alpha } \)  includes the effects of magnetization dissipation via Gilbert damping. When magnetization dissipation does not occur,  \( g_{ \alpha } \)  vanishes. In this case, the total resistance of the device is the sum of the resistances of the left and right metal-magnetic insulator contacts. This result implies that there is no resistance in the bulk of the magnetic insulators, and spin transport is dissipationless therein.  \( g_{ \alpha } \)  is proportional to the Gilbert damping coefficient and the length of the device, i.e., the total amount of dissipation versus the injection and detection efficiencies. A signature of spin supercurrent is, therefore, that it decays algebraically as a function of the length of the magnetic insulator \cite{138takei}. This phenomenon contrasts with the expectations of diffusive spin transport via magnons that decay exponentially when the system exceeds the spin relaxation length \cite{2cornelissen,139cornelissen}.

However, in ferromagnets, the simple picture above for the spin-injection and spin-detection in a ferromagnet where the current is governed by superflow is incomplete. In such setups, the ubiquitous dipole-dipole interaction dramatically affects spin transport and reduces the range to a few hundred nanometers even in ferromagnetic films as thin as five nanometers \cite{140brataas}. In antiferromagnets, long-range dipole-dipole interactions do not occur. Spin superfluidity is also possible in AFIs, but the first predictions for the range of the spin supercurrent were short and less than microns \cite{141takei}. In biaxial antiferromagnets, the anisotropy hinders superfluidity by creating a substantial threshold that the current must overcome. Nevertheless, the application of a magnetic field removes this obstacle near the spin-flop transition of the antiferromagnet. Spin superfluidity can then persist across many microns \cite{142qaiuzadeh}.

Thus far, in our opinion, electrical control of magnon condensation and spin superfluidity has not yet been experimentally realized. A recent report claims signatures of a long-distance spin supercurrent through an antiferromagnet \cite{143yuan}. However, thermal transport is the basis of the results, crucially without detecting the expected accompanying spin-injected transport signal. As discussed above, other effects, for instance, spatially-extended temperature gradients can explain nonlocal signals.

There are reports of supercurrents in other geometries using thin ferromagnetic layers of YIG. Ref.\ \cite{Bozhko:NatPhys2016} first created a Bose-Einstein condensate of magnons by parametric pumping. Additionally, they heated the condensate in a spatially confined region and measured the spatiotemporal decays of the condensate after turning the pump off. They found that the condensate decays faster in the hot region when there is a thermal gradient in the system. The interpretation is that there is a current from the hot condensate to the cold condensate that speeds up the decay in the hot area. This additional decay channel represents an indirect observation of a spin supercurrent in a magnonic system. 

More complicated ways of transferred spin information are possible in condensates. Second sound is a propagation of the properties of the condensate. Ref.\ \cite{Tiberkevich:SciRep2019} reports the excitation of coherent second sound. Ref.\ \cite{Bozhko:NatCom2019} reports the observation of Bogoliubov waves. Humps in the condensate density propel many hundreds of micrometers out of heated areas.

\subsection{Magnon-induced Superconductivity}

In metals, superconductivity arises from an effective attractive electron-electron interaction. In low temperature superconductors, phonons mediate the attraction between the electrons via the electron-phonon interaction. The interaction is attractive for electrons with energies roughly within the Debye energy of the Fermi surface. This leads to a Cooper instability, the formation of Cooper pairs, and a phase transition from a normal state to a superconducting state. In the superconducting state, charge transport is dissipationless and the Meissner effect causes a perfect diamagnetic screening of external magnetic fields. 

In high-temperature superconductors, the superconducting phase is typically close to the antiferromagnetic phase. It is then likely that the effective attraction between the electrons arises from or is related to the spin fluctuations \cite{Moriya2000ap}. However, a widely accepted and established theory of high-temperature superconductors is yet to be established. The superconducting gap is a spin-singlet with a d-wave orbital symmetry. There are also materials that exhibit superconductivity with more exotic symmetries. 

Similar to phonons, magnons are bosons. Magnons also interact with electrons via the electron-magnon interaction. In the systems of the present interest, magnetic insulators in contact with normal metals, the electron-magnon interaction is at the metal-insulator interface. It is then natural to ask if magnons can cause superconductivity in layered systems of magnetic insulators and conducting materials. The advantage of such systems is that the properties of the electrons, the magnons, and the electron-magnon interaction can be engineered by selecting different materials with desired bulk and interface characteristics. In this way, the magnons and the electrons can be independently tuned. This opens a new door towards exploring superconducting properties, possibly with unprecedented properties. The exploration of such systems is therefore desirable. As an added benefit, unraveling these mysteries could perhaps also give additional insight into the mechanisms behind high-temperature superconductors. 

Kargarian et al.\ theoretically explored the possible pairing induced by magnons at the surface of topological insulators \cite{Kargarian:PRL2016}. They considered the surface of a doped 3D topological insulator in contact with a ferromagnetic insulator. In topological insulators, the electron states are helical and spin and momentum are locked.  Because of the unusual electronic states, the pairing also becomes unconventional. Ref.\ \cite{Kargarian:PRL2016} finds that the effective interaction between the electrons is rare. It is attractive between electrons near the Fermi surface with the same momentum. They the resulting pairing is called Amperian pairing. 

There are experimental observations of superconductivity that breaks time-reversal symmetry in epitaxialy bilayer films of bismuth and nickel \cite{Gong:ScienceAdvances:2017}. The demonstration is the onset of the polar Kerr effect when the system becomes superconducting. The critical temperature is 4K. The results may arise from magnetic fluctuations in nickel. The strong spin-orbit coupling and lack of inversion symmetry can induce an exotic pairing. 

Independent of, and before these developments and observations were reported, but likely after they were initiated, one of the present authors started exploring the possibility of magnon-induced superconductivity in thin normal metal films in contact with ferromagnetic and antiferromagnetic insulators. Motivated by the successful demonstration of spin-pumping and spin-transfer torques in bilayers of ferromagnetic insulators and normal metals, the question was if the same systems also could give rise to magnon-induced superconductivity. The combination of the well-explored features of the electron-magnon coupling across interfaces in such systems with the possibility of pairing therein gives a rich playground for exploring new superconducting states. We will now outline the simplest version of the theory of magnon-induced superconductivity in such system, hoping to motivate the first experimental detection of such a fascinating and underexplored phenomena.

Let us consider a two-dimensional metal in contact with ferromagnetic insulators. In the absence of interactions between the two subsystems, the  electrons are spin-degenerate with a Hamiltonian 
\begin{equation}
    H_e = \int d {\bf q} \epsilon_q c^\dag_{{\bf q}s} c_{{\bf q}s} \, ,
\end{equation}
where $\epsilon_q$ is the electron eigenenergy, ${\bf q}$ is the momentum, $s$ is the spin, and $c_{{\bf q}s}$ annihilates an electron with momentum ${\bf q}$ and spin $s$. The Hamiltonian of the magnons in the left ferromagnetic insulator is 
\begin{equation}
    H_{ml} = \int d {\bf k} \hbar \omega_{\bf k} l^\dag_{{\bf k}} l_{{\bf k}}
    \label{eq:Hlm}
\end{equation}
where $\hbar \omega_{\bf k}$ is the magnon energy and $l_{\bf k}$ annihilates a magnon with momentum ${\bf k}$. Similarly, the Hamiltonian of the magnons in the right ferromagnetic insulators is represented by the Hamiltonian $H_{rm}$ as in Eq.\ (\ref{eq:Hlm}) with $l_{\bf k} \rightarrow r_{\bf k}$, where $r_{\bf k}$ annihilates a magnon with momentum ${\bf k}$ in the right ferromagnetic insulator. 

It is well established that localized spins in ferromagnetic insulators can interact with itinerant spins in adjacent conductors. The simplest effect of the electron-magnon interaction is to induce effective Zeeman fields in the normal metal. Such spin splittings can be detrimental to superconductivity when they are large enough. To simplify the discussions and avoid such complications, we assume that the left and right magnetic insulators are identical, but that the magnetizations are anti-parallel. The induced Zeeman fields from the left and right magnetic insulators then cancel each other. In this regime, the interaction is especially simple and determined by $H_{i}=H_{il}+H_{ir}$, where e.g. the interaction between the electrons and the magnons in the left ferromagnet is
\begin{equation}
    H_{il} = \int d {\bf q} \int d {\bf k} V \left[ a_{\bf k} c_{{\bf q}+{\bf k}\downarrow}^{\dag} c_{{\bf q} \uparrow} + \mbox{h.c.} \right] \, , 
    \label{eq:Heml}
\end{equation}
where $V$ is the momentum-representation of the interface exchange coupling strength between the electrons in the normal metal and the magnons in the left magnetic insulator. In terms of the real-space microscopic interface coupling between the conduction electrons and the localized spins $J_I$, $V=-2J_I \sqrt{s/2N}$, where $s$ is the spin quantum number of the localized spins and $N$ is the number of localized spins \cite{Rohling:PRB2018}.  The expression describing the interaction between the electrons in the normal metal and the magnons in the right magnetic insulator is similar to Eq.\ (\ref{eq:Heml}). Taking both interfaces into account, to second order in the electron-magnon coupling strength $V$ and for electron pairs with opposite momenta, the effective electron-electron interaction becomes
\begin{equation}
    H_{ee} = \sum_{{\bf k}{\bf q}} V_{{\bf k}{\bf q}} c_{{\bf k}\downarrow}^\dag c_{-{\bf k} \uparrow}^\dag c_{-{\bf q} \uparrow} c_{{\bf q}\downarrow} \, , 
    \label{eq:Hee}
\end{equation}
where the strength 
\begin{equation}
    V_{{\bf k}{\bf q}} = 2 V^2 \frac{\hbar \omega_{{\bf k}+{\bf q}}}{(\hbar \omega_{{\bf k}+{\bf q}})^2-(\epsilon_{\bf k}-\epsilon_{\bf q})^2}
\end{equation}
is determined by the quasi-particle energies of the magnons and electrons, as well as the interface exchange coupling strength between the electrons and the magnons.

In the mean-field approximation, the superconducting properties are determined by the superconducting gap. The gap can be a spin singlet ($S=0$) or a spin triplet ($S=1$). When the gap is a spin singlet, it is an even function of the momentum. For spin triplet gaps, the gap is an odd function of momentum. In ferromagnetic insulator-metal-ferromagnetic insulator systems, the effective electron-electron interaction of Eq.\ (\ref{eq:Hee}) causes pairing between electrons with opposite spins. Consequently, for both spin singlet and spin triplet states, the total spin of the Cooper pair is zero, $S_z=0$.

We consider the gap at the critical temperature, $T=T_c$. For spin triplet pairing, the gap equation becomes \cite{Rohling:PRB2018,Fjaerbu:PRB2019,Erlandsen:PRB2019}
\begin{equation}
    \Delta_{{\bf k}T} = - \sum_{{\bf q}} V_{{\bf k}qO} \Delta_{{\bf q}T} \chi_q \, , 
    \label{eq:gap}
\end{equation}
where $\chi_q=\tanh{(|\epsilon_{\bf q}|/2k_B T)}/(2 |\epsilon_{{\bf q}}|)$ and $V_{{\bf k}qO}=( V_{{\bf k}{\bf q}} -V_{-{\bf k} {\bf q}})/2$ is odd (O) in the momentum ${\bf k}$. There is a similar equation for spin singlet pairing described by the gap $\Delta_{{\bf k}s}$, where $V_{{\bf k}qO} \rightarrow V_{{\bf k}qE}$ and $V_{{\bf k}qE}=( V_{{\bf k}{\bf q}} +V_{-{\bf k} {\bf q}})/2$ is even (E) in the momentum ${\bf k}$. 

In the gap equation (\ref{eq:gap}), at the Fermi surface, $\epsilon_{\bf k}=\epsilon_{\bf q}=\epsilon_F$, where $\epsilon_F$ is the Fermi energy. Both when the momenta ${\bf k}$ and ${\bf q}$ are parallel  (${\bf k}-{\bf q}$ is small) and when they are anti-parallel (${\bf k}+{\bf q}$ is small), the even combination of the effective interaction, $V_{{\bf k}qE}$, is positive and repulsive. The system will therefore not favor spin singlet supercondutivity. Instead, spin triplet superconductivity is preferred. When the momenta ${\bf k}$ and ${\bf q}$ are parallel  (${\bf k}-{\bf q}$ is small), the odd combination of the effective interaction, $V_{{\bf k}qO}$, is negative and attractive. In contrast, when the momenta ${\bf k}$ and ${\bf q}$ are anti-parallel  (${\bf k}+{\bf q}$ is small), $V_{{\bf k}qO}$, is positive and this is compensated by the fact that $\Delta_{-{\bf q}T}=-\Delta_{{\bf q}T}$. As well established, in ferromagnets, magnons induced triplet superconductivity \cite{Karchev:EPL2015}.

We can give estimates of the critical temperature based on the many studies of spin-transfer torques, spin-pumping and induced exchange fields in ferromagnetic insulators-normal metal systems. For instance, in the case of the ferromagnetic insulator YIG, we can base the estimates of the interface exchange coupling $J_I$ on the measured spin-mixing conductance \cite{Heinrich:PRL2011,75burrowes}. Whereas for the ferromagnetic insulator EuO, we can use the measurements of an induced Zeeman field \cite{Tkaczyk:1988}. For YIG-Au-YIG trilayers, it is found that the critical temperatures range between 0.5K and 10K. Similarly, for EuO-Au-EuO trilayers, it is found that the critical temperature is in the range 0.01K to 0.4K. Computations of the critical temperature are extremely sensitive to the values of the of the interface exchange interaction. The uncertainty in the extraction of this crucial parameter from experiments should motivate direct experimental exploration of the possible intriguing behavoir of magnon-induced superconductivity in layered systems.

As in ferromagnets, magnons in antiferromagnetic insulator also couple strongly to conduction electrons in adjacent metals. This interfacial tie can also lead to magnon-induced superconductivity in systems consisting of metals and antiferromagnetic insulators. The strong exchange interaction causes the resonance frequencies to be much higher in antiferromagnets than in ferromagnets. In simple antiferromagnets, there are two sublattice with antiparallel magnetic moments at equilibrium. There are therefore two kinds of magnons. The ratio between the coupling of these magnons to the electrons depends on the microscopic properties at the interface. In the simplest case, at compensated interfaces, the spins at the two sublattices couple equally strong to the itinerant electrons. In uncompensated interfaces, there is a stronger coupling to the localized spins in one sublattice with respect to the coupling to the localized spin in the other sublattice. 

In tight-binding models, when there is a perfect lattice matching between the conduction band sites in the normal metal and the sites of the localized spins in the antiferromagnet, Umklapp scattering is important \cite{69cheng,141takei}. In antiferromagnets, the analogous electron-magnon coupling to Eq.\ (\ref{eq:Heml}) then involves the Umklapp momentum. At half-filling, an Umklapp process takes a state at the Fermi surface to another state at the Fermi surface \cite{Fjaerbu:PRB2019}. This feature influences the pairing interaction. As a consequence, at half-filling, the pairing is a spin-singlet with d-wave orbital symmetry \cite{Fjaerbu:PRB2019}. The estimates are that the critical temperature in MnF$_2$-Au-MnF$_2$ is of the order of 1K.

Away from half-filling, the strength of the effective electron-electron interaction is similar to the ferromagnetic case described by Eq.\ (\ref{eq:Hee}), but with a renormalized strength
\begin{equation}
V_{{\bf k}{\bf q}}^{(A)} = V_{{\bf k}{\bf q}} A({\bf k}+{\bf q},\Omega)
\label{eq:VkpAF}
\end{equation}
where the factor $A$ can be understood as arising from the constructive or destructive interference of squeezed magnons \cite{Erlandsen:PRB2019}. Similar to ferromagnets, the electron-magnon interaction of Eq.\ (\ref{eq:VkpAF}) gives rise to spin-triplet pairing. 

In quantum mechanics, the uncertainty of two canonically conjugate variables cannot vanish at the same time. Nevertheless, squeezed states exists where the uncertainty in one variable is reduced while the uncertainty in the other variable is enhanced. In antiferromagnets, one physical picture is that the magnons that diagonalize the spin Hamiltonian result from squeezing \cite{Kamra:PRB2019}. The elementary excitations are squeezed magnon states. Unlike the magnons in the initial basis, the squeezed magnon states contain an average spin on each sublattice that is much larger than the unit net spin \cite{Kamra:PRB2019}. Yes, at the same time, the compensation between the spins at the two sub-lattices implies that only one unit spin is excited. These features have important implications for the physics when antiferromagnets are coupled to normal metal. 
At interfaces, when the conduction electrons couple to only one of the sublattices, the spins in the normal metal effectively couples to a large spin in the antiferromagnetic insulator. As a result, the factor $A$ in Eq.\ (\ref{eq:VkpAF}) is greatly enhanced because the squeezed magnon carries a large at the interface. Therefore, the electron-magnon interaction becomes much stronger \cite{Erlandsen:PRB2019} at uncompensated interfaces as compared to interfaces that couple in similar ways to both sub-lattice. This effect can significantly enhance the coupling between the electrons and the magnons. In turn, the estimates of the critical temperature then significantly exceeds 1K.

We encourage experimental explorations of the possibility of observing superconductivity in thin metallic films sandwiched between magnetic insulators. It might become possible to engineer such systems with controlled superconducting properties at measurable temperatures.

\section{Conclusions}

We present in this review recent developments of the utilization of spins in magnetic insulators to control electric signals. In magnetic insulators, the lack of mobile charge carriers often implies a considerably longer spin coherence time as compared to metallic systems. Both ferromagnetic and antiferromagnetic insulators are of interest. While the former materials are quite widely studied, the latter systems have the advantages of faster response time, are less explored and features intriguing quantum aspects.  In the Terahertz gap, there are no state-of-the-art technologies for generating and detecting radiation. Using antiferromagnet insulators, in combination with metals, can enable the development of Terahertz electronics devices. 

We have seen that spins can propagate across micrometers in a wide range of magnetic insulators. This long spin coherence opens the possibility of exploring the fascinating phenomena of electrical control of magnon condensation and spin superfluidity. Such control enables the use of the magnon phase-coherence and will reveal details of the magnon-magnon and magnon-phonon interactions, also in confined systems. 

At low temperatures, magnons might even mediate superconductivity in adjacent conductors. Such systems are of interest since the superconductivity might be unconventional. Furthermore, new ways of forming thin metallic systems or purely two-dimensional systems can open up new doors for controlling the electronic and magnonic properties, possibly with a larger critical temperature as a result.

\section{Acknowledgements}

A.\ B.\ would like to thank Eirik Løhaugen Fjærbu, Niklas Rohling, Akashdeep Kamra, Øyvind Johansen, Rembert Duine, Eirik Erlandsen, and Asle Sudbø for stimulating discussions and collaborations. 
A.\ B.\ has received funding from the European Research Council via Advanced Grant No. 669442 "Insulatronics" and the Research Council of Norway Grant through its Center of Excellence funding scheme, Project No. 262633 "QuSpin".

\section{Declarations of Interest}

None.

\bibliography{bibliography}

\end{document}